\documentclass[conference]{IEEEtran}
\IEEEoverridecommandlockouts
\usepackage{cite}
\usepackage{amsmath,amssymb,amsfonts, amsthm}
\usepackage{graphicx}
\usepackage{textcomp}
\usepackage{mathtools}
\usepackage{color,soul}
\usepackage{algorithm}
\usepackage{algpseudocode}
\usepackage[most]{tcolorbox}

\usepackage{algorithm}
\usepackage{algpseudocode}
\usepackage{hyperref}
\usepackage{orcidlink}

\usepackage{booktabs}
\usepackage{multirow}

\definecolor{rqbg}{RGB}{232,238,255} % tweak to taste
\definecolor{revblue}{RGB}{0,140,255}

\newcommand{\rqsummarybox}[1]{%
  \par\medskip
  \noindent
  \begingroup
  \setlength{\fboxsep}{6pt}% inner padding
  \colorbox{blue!10}{%
    \parbox{\dimexpr\linewidth-2\fboxsep\relax}{#1}%
  }%
  \endgroup
  \par\medskip
}

\def\BibTeX{{\rm B\kern-.05em{\sc i\kern-.025em b}\kern-.08em
    T\kern-.1667em\lower.7ex\hbox{E}\kern-.125emX}}
\begin{document}

\title{Rings Around the Ancilla: A Workload-Aware FTQC Architecture
}

\author{
\IEEEauthorblockN{
Archisman Ghosh\IEEEauthorrefmark{1}\orcidlink{0000-0002-0264-6687},
Avimita Chatterjee\IEEEauthorrefmark{1}\IEEEauthorrefmark{2}\orcidlink{0009-0001-7421-9334},
Swaroop Ghosh\IEEEauthorrefmark{3}\orcidlink{0000-0001-8753-490X}
}

\IEEEauthorblockA{\IEEEauthorrefmark{1}\textit{Department of Computer Science and Engineering}, The Pennsylvania State University}
\IEEEauthorblockA{\IEEEauthorrefmark{2}\textit{Lawrence Berkeley National Laboratory}}
\IEEEauthorblockA{\IEEEauthorrefmark{3}\textit{School of Electrical Engineering and Computer Science}, The Pennsylvania State University}
\IEEEauthorblockA{Emails: \{apg6127, szg212\}@psu.edu, AvimitaChatterjee@lbl.gov}
}
% \and
% \and
% \and

\maketitle

\begin{abstract}
Obtaining a quantum advantage in practical applications over classical machines requires a fault-tolerant model of computation. However, the resource overhead of developing a fault-tolerant quantum computer (FTQC) architecture is quite high. Current state-of-the-art architecture designs promise constant-time access to logical qubits at the expense of a high qubit overhead, or qubit-dense layouts at the expense of increased latency in the execution of a quantum workload. In this paper, we propose (a) an architecture design that balances the tradeoff of execution time of workloads with the density of logical qubits by placing the surface code patches around the ancilla region, giving almost equal access of the ancilla region to all the data qubits, (b) design a novel algorithm to leverage the $T$-gate characteristics of the user workload to suggest the optimal placement strategy in the floorplan of our architecture, (c) a workload-selective fast-$Y$ optimization knob that accelerates $Y$-basis measurements only for qubits whose $Y$-gate demand justifies the additional tile cost, and (d) explore the scope of executing multiple programs concurrently, which enhances the reconfigurability of our proposed design. Through numerical evaluation, we observe that the proposed architecture design maintains the cycles per instruction time close to optimal while using up to $\sim21\%$ fewer data tiles, and up to $\sim90\%$ efficiency under a 10-way concurrency.
\end{abstract}

\begin{IEEEkeywords}
Fault-Tolerant Quantum Computing, Quantum Computing Architecture, Surface Code, Quantum Error Correction
\end{IEEEkeywords}

\section{Introduction}
Fault-tolerant quantum computation (FTQC) is widely regarded as the first realistic path to practical quantum advantage, but it demands orders-of-magnitude more qubits and time than the capability of noisy intermediate-scale devices. Quantum error-correcting codes such as the surface code protect logical qubits by spreading information over a two-dimensional lattice, transforming algorithms into long-lived patterns of patches, stabilizer measurements, lattice surgery, and magic-state injections. At this scale, performance is no longer determined solely by the abstract circuit depth, but by the \emph{physical realization} of that circuit on a finite 2D fabric.
We refer to this physical realization as the hardware ``floorplan”: the placement and sizing of logical data tiles, ancilla resources for syndrome extraction and multi-qubit measurements, routing corridors for deformation or transport, and the allocation of magic-state factories and their injection interfaces. The objective is to minimize the end-to-end space–time cost, subject to hardware and code constraints, while sustaining sufficient measurement bandwidth and avoiding congestion. Poorly designed floorplans can limit algorithmic speedups through increased routing latency, contention for ancilla regions and magic state factories, and/or underutilized qubits due to poor allocation. Therefore, it is in our best interest to treat FTQC floorplanning as a coupled space-time optimization problem rather than separate optimizations for latency and area. Prior FTQC floorplanning efforts have neglected the idea of treating these objectives jointly, and also not taken into consideration the workload structure and the scope of concurrent execution.

\begin{figure*}    
    \centering
    \includegraphics[width=1\linewidth]{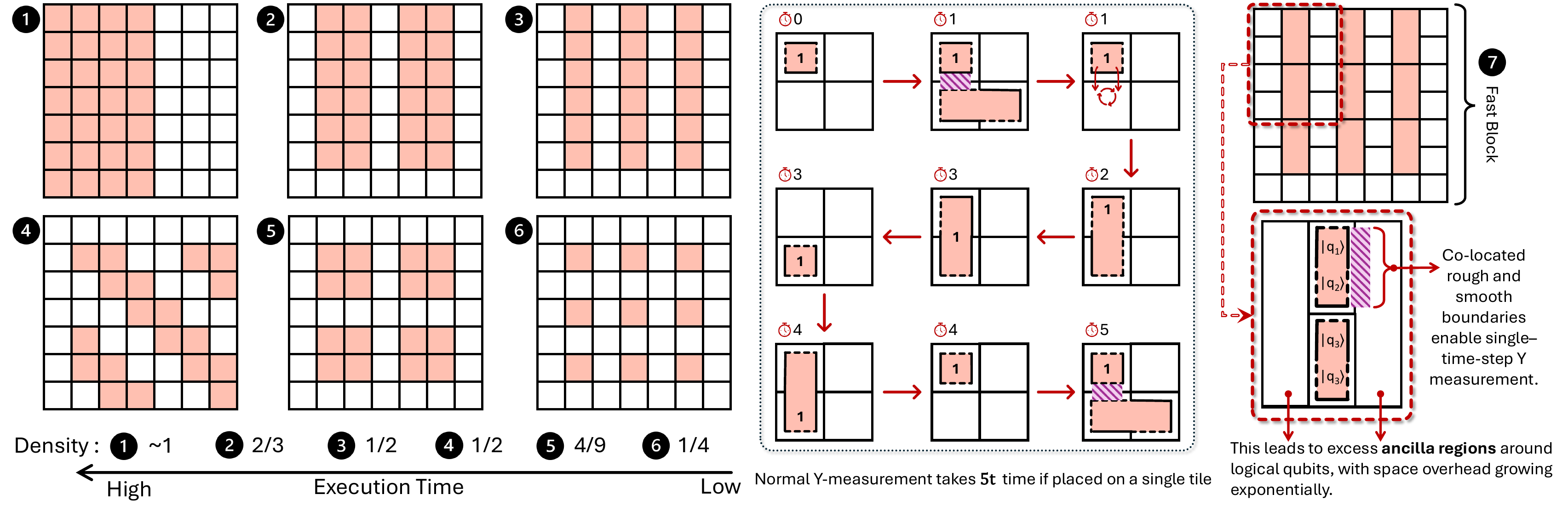}
    %\vspace{-10pt}
    \caption{\textbf{Prior research in surface-code floorplan design.} Designs 1 \cite{lsqca}, 2 \cite{Lee2021TensorHypercontraction}, 3 \cite{beverland2022assessingrequirementsscalepractical}, 4 \cite{fujiu2025densepackingsurfacecode}, 5 \cite{Chamberland2022TwistFreeTemporallyEncoded}, and 6 \cite{Beverland_2022} are arranged in decreasing order of data tile density. We find that the sparser the floorplan design, the faster the execution at the expense of a higher qubit overhead. In design 7 \cite{Litinski_2019}, we see that the $Y$-basis measurement (which takes $5t$ when logical qubits are represented by a single tile) takes $1t$ to execute due to the presence of both the $X$ and $Z$ stabilizers on the same edge, but at a huge tile overhead.}
    \label{fig:intro1}
    % \vspace{-10pt}
\end{figure*}

\textbf{Lack of joint optimization:} Most existing FTQC architectures optimize the floorplan along a single axis, viz., time and space. On the \emph{time} axis, there has been research that focuses on developing a floorplan design that guarantees almost constant-latency access to arbitrary pairs of logical qubits by reserving substantial auxiliary area for lattice surgery and routing. For example, we observe from Fig. \ref{fig:intro1} that in \cite{Beverland_2022}\cite{Chamberland2022TwistFreeTemporallyEncoded}\cite{beverland2022assessingrequirementsscalepractical, fujiu2025densepackingsurfacecode}\cite{Lee2021TensorHypercontraction}, the authors have designed floorplans with a data tile density of $1/4$, $4/9$, $1/2$, and $2/3$, respectively. Although this allows the execution of arbitrary operations in a constant time, a lot of the space is reserved for ancilla qubits that reduces the data-tile density of the floorplan and adds to the qubit budget of the hardware design. Given a 400-logical-qubit workload, a $50\%$-density floorplan immediately implies a fabric budget of $\ge 800$ logical tiles even before accounting for factories and routing corridors. Since each logical tile is realized as a distance-$d$ surface-code patch requiring $\Theta(d^2)$ physical qubits, this dilution translates directly into a quadratic amplification of the physical-qubit budget, substantially inflating the overall system overhead. On the \emph{space} axis, the premise of having data tiles close together on the floorplan reduces the congestion and resource contention among the ancilla tiles, thus increasing the density of data tiles. Authors in \cite{lsqca} show that investing floorplan area to push for a constant-time execution of workloads is not a good design choice due to system-level delays like generating magic states based on standard protocols \cite{Litinski_2019_msd}. However, on the contrary, we observe that having a high data tile density usually contributes to a longer or variable-latency access due to movement in the data-dense region. In $T$-gate heavy workloads, this latency adds up, leading to delayed execution, which is not desirable while achieving near-term fault tolerance. For example, a stark increase in the CPI (code beat per instruction) in the arithmetic circuits is observed in \cite{lsqca}. Therefore, optimizing ``time" on a uniform substrate overlooks that the floorplan constrains physically realizable parallelism, while optimizing ``space" under generic-access assumptions ignores real workload locality, hotspots, and phase behavior. The natural conclusion from the above observation is that the space-time cost of FTQC should be optimized \emph{jointly}, which motivates our approach: we explore the two axes simultaneously by coupling a physically grounded floorplan with workload-aware qubit placement and movement/scheduling, targeting the Pareto frontier between (i) high data tile density and (ii) low end-to-end latency under realistic bottlenecks. 

\textbf{Existing placement abstractions neglect workload structure and gate-dependent costs:} The cost of implementing non-Clifford gates on a surface code architecture is huge in terms of resource requirements and magic-state factory throughput \cite{Gidney_2024}. Current floorplan-level placement policies attempt to optimize the mapping and routing of the surface code patches by formulating the problem as a 3D-path routing problem \cite{hamada2026efficienthighperformanceroutinglatticesurgery} and solve it using dependency graphs \cite{dascot}\cite{silva} and edge-disjoint paths \cite{Beverland2022EdgeDisjointPaths}. Specific $T$ gates like the $Y$ gates take more surface code cycles ($5t$ instead of $1t$ for $X$ or $Z$-basis measurement when using one ancilla tile, $t$ being a surface code cycle) to get measured due to the rotation of patches in the floorplan. As an attempt to reduce the latency overhead in $Y$-basis measurements, the \emph{Fast} block was designed in \cite{Litinski_2019} by keeping the $X$ and $Z$ stabilizers on the same edge of the extended data tile (Fig. \ref{fig:intro1} (7)). However, this has a huge resource overhead of the order $O(n+\sqrt{n})$ data tiles for representing $n$ logical qubits. Moreover, these abstractions do not take the cost of implementing a Non-Clifford gate into account while placing them on the floorplan. Designs like the \emph{Fast} block keep a similar structure of the tiles for all qubits uniformly, which leads to a huge wastage of ancilla to achieve execution in a single surface code cycle. Therefore, there is a deterministic trade-off between enabling a fast $Y$ measurement to reduce latency at the cost of extra qubits. The goal is to re-introduce it as a selective optimization knob in the architecture design, to enable the qubits that are heavy in $Y$ measurements to be exposed to the fast-$Y$ measurement and hence motivate a workload-aware policy that promotes the qubits whose $Y$-measurement latency exceeds the placement cost.

\textbf{Lack of support for concurrency:} While multiprogramming has been widely explored for NISQ systems, direct studies of concurrent execution of multiple independent workloads on shared FTQC floorplans remain limited. Existing FTQC work has only recently begun to consider settings such as \emph{online job arrival} on lattice-surgery-based architectures \cite{Wakizaka_2025}. Following emerging trends motivate the need to study concurrency in the domain of FTQC: (a) the diminishing physical error rates which will necessitate lower distance surface codes reducing the overhead of error correction, (b) emergence of applications such as quantum machine learning which may work well with partial FTQC, (c) current progress in low overhead fault-tolerant architectures and (d) public roadmaps and vendors claim of large-scale quantum hardware with specific qubit technologies \cite{ibm_ftqc_2025, microsoft_darpa_2024, microsoft_majorana1_2025}  \cite{atom_ac1000_2026, atom_ft_milestone_2023, atom_preview_2026}. The above factors will either reduce the overhead of FTQC, freeing up resources, and/or enable backends with a large number of qubits, likely by the end of the decade, which in turn can benefit from concurrency. Although seemingly simple at first, concurrency may bring new challenges, such as sharing resources fairly among the independent workloads and scheduling the shared resources efficiently. This warrants the inclusion of this important consideration in the FTQC architecture design.

\begin{figure}    
    \centering
    \includegraphics[width=1\linewidth]{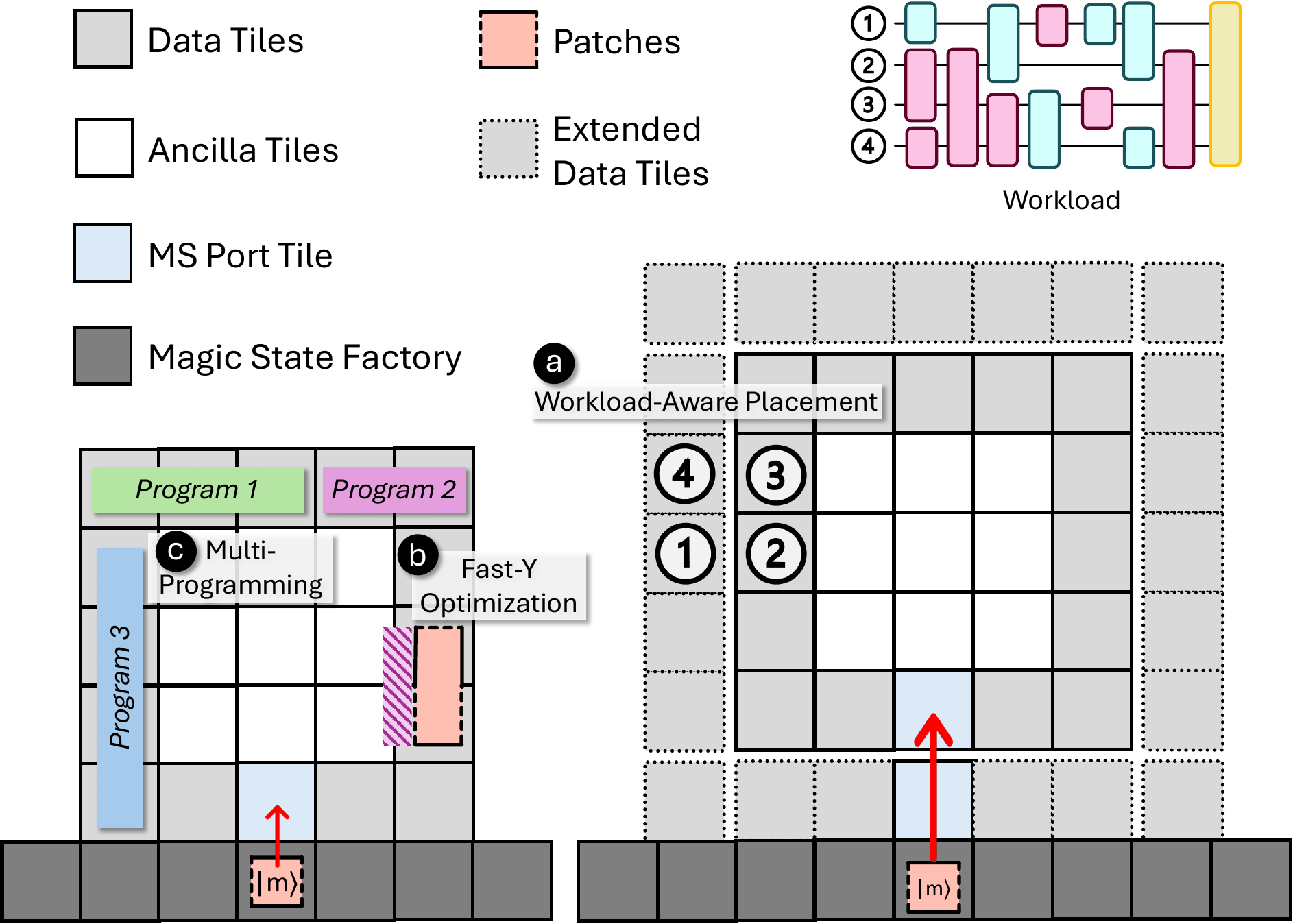}
    %\vspace{-10pt}
    \caption{\textbf{The proposed architecture.} The first structure represents the design of the baseline architecture with the data tiles in the outer ring and the ancilla tiles in the centrally located computational region. The second structure describes the extension of the geometry using data tiles that are stacked on the outside. We show a single entry point to the CR in this diagrammatic representation, which is used by the magic state factory to aid in the measurement of $T$ gates. In (a) we represent the workload-aware placement, (b) we represent the fast-$Y$ optimization knob, and (c) we represent the scope of multiprogramming. The magic state factory is diagrammatically represented by a single row but occupies as many tiles as it needs for producing high-fidelity magic states.} 
    \label{fig:arch}
    \vspace{-10pt}
\end{figure}

\textbf{Proposed idea:} We propose a stacked-annulus layout (Fig. \ref{fig:arch}) to organize the data tiles (with surface code patches to correct the logical qubit represented by the data tile) into concentric data rings wrapped around a central compute region (comprising the ancilla tiles). This approach exposes short and nearly uniform access from the data tiles that are innermost to ancilla while allowing outer tiles to host the qubits with low computation overhead. This way, we tackle the first problem by exploring in both directions of optimization, viz., data tile density as well as execution time. In the proposed design, the data-tile density is preserved to some extent while providing an almost constant-time execution for FTQC workloads. We discuss the details of the architecture design, the movement primitives, and their extension for larger workloads (Fig. \ref{fig:arch}(a)) in Section \ref{sec:archdesign}. To tackle the second problem, we exploit the $T$-gate density of a qubit in a workload along with the multi-qubit interaction degree to search for an optimal placement pattern that reduces the latency incurred to move the data tiles around the floorplan in our design, which is discussed in Section \ref{sec:placementandoptim}. In addition, we also introduce the $Y$-basis measurement in the \emph{Fast} block \cite{Litinski_2019} as a reprogrammable optimization knob (Fig. \ref{fig:arch}(b)). We find that selectively choosing the qubits with a higher $Y$-gate density (single- and multi-qubit) reduces the total qubit count and justifies the use of one extra tile for representing the qubit to provide a net gain in the overall execution time. While dealing with concurrent independent workloads, a naive policy of compartmentalizing the floorplan to reserve space for the workloads is not sufficient. We explore placement policies and the $Y$-measurement optimization knob on the proposed floorplan design, and determine the overall performance and throughput in the presence of multiple independent workloads (Fig. \ref{fig:arch}(c)) in Section \ref{sec:sharing}. We numerically evaluate the proposed approach on a wide set of benchmarks that cater to near-term FTQC, including a custom-generated set of circuits with varying $T$-gate densities, along with Hamiltonian simulations (for H$_2$O, H$_2$, and LiH molecules), arithmetic circuits (adder, multiplier, and square root), and Quantum Fourier Transform (QFT) circuits in Section \ref{sec:eval}, discuss related work in Section \ref{sec:related}, and conclude the paper in Section \ref{sec:conc}. \emph{To the best of our knowledge, this is the first attempt at designing an FTQC surface-code architecture that takes into account the workload characteristics during qubit placement and explores concurrent execution of workloads, all while maintaining a decent tradeoff between data tile density and overall execution time.}

\section{Background}
This section describes the basics of surface code and lattice surgery, and the basics of FTQC architecture based on the Clifford+$T$ gate set.
\textbf{Surface-Code Architecture Model: }
We assume a fault-tolerant architecture based on the planar surface code \cite{Litinski_2018}\cite{Litinski_2019}\cite{ghosh2025designautomationquantumerror}, in which logical qubits are encoded as 2-D patches of physical qubits laid out on a local nearest-neighbor fabric. A distance-$d$ logical patch occupies $O(d^2)$ physical qubits and is protected by repeated rounds of stabilizer measurement. Each patch has two $X$-type and two $Z$-type boundaries, so that strings of physical Pauli operators connecting like boundaries realize the logical operators $\bar X$ and $\bar Z$. In this work, we use the standard patch-based abstraction of the surface code, where the unit of computation is not an individual physical qubit but a logical patch together with its geometric boundaries.
Logical interactions are implemented using lattice surgery, which preserves the locality constraints of the underlying two-dimensional architecture \cite{Chatterjee_2025}\cite{dascot}\cite{Horsman_2012}. 
At the architectural level, we model the floorplan as a grid of logical tiles,
\(
\mathcal{T}=\{T_{i,j}\},
\)
where each tile denotes a fixed surface-code footprint at distance $d$. This tile-based abstraction allows the physical floorplan to be described at the logical level while retaining the locality constraints imposed by the surface code.
We distinguish three architectural resources. First, \emph{data patches} are long-lived logical patches that store the logical qubits. Their geometric arrangement determines which joint Pauli measurements can be realized directly through local lattice surgery and, consequently, the movement or reshaping required before interaction. Second, \emph{ancilla} or \emph{compute} regions provide temporary workspace for syndrome extraction and lattice surgery operations. Third, \emph{magic-state distillation} resources, organized as dedicated distillation blocks and factories that inject high-fidelity magic states to implement the non-Clifford gates\cite{Litinski_2019_msd}\cite{Bravyi_2012}. 
% \begin{figure}    
%     \centering
%     \includegraphics[width=1\linewidth]{samples/fig/surface_code.pdf}
%     %\vspace{-10pt}
%     \caption{An abstraction of the surface code architecture. The orange patches are the surface codes and the white tiles are the ancilla in the floorplan.\hl{Don't see the point of this fig. It doesn't explain any issue or present any new info. We have bigger problems that need explaining and showing in intro. Delete.}}
%     \label{fig:surfacecode}
%     % \vspace{-10pt}
% \end{figure}

% Fig. \ref{fig:surfacecode} illustrates this abstraction at two levels. At the tile level, a logical patch occupies one or more tiles within the architectural grid. At the physical level, each tile corresponds to a surface-code patch with explicit boundaries and stabilizer structure. This abstraction is the basis for the floorplanning, placement, and movement policies developed in the remainder of the paper.
\textbf{Clifford+$T$ Execution Model: }
We assume input circuits over the universal Clifford+$T$ gate set
\(
\{H,S,\mathrm{CNOT},T\}.
\)
Following standard surface-code compilation~\cite{Litinski_2019}, the circuit is rewritten as Pauli-product rotations and measurements: Clifford gates become \(\pi/4\) rotations and \(T\) gates become \(\pi/8\) rotations. By commuting all Clifford rotations to the end and absorbing the final Clifford block into the measurement layer, we obtain an equivalent circuit consisting of \(\pi/8\) Pauli-product rotations followed by terminal Pauli measurements. Consequently, the dominant cost arises from the non-Clifford rotations, which require magic-state consumption, while Clifford corrections are tracked in the Pauli frame.

\section{Proposed Floorplan design}\label{sec:archdesign}

\subsection{Overview}

The baseline architecture is an annulus of data tiles wrapped around a central compute region (CR), with a set of tiles reserved for magic-state factories (MSFs), placed below it. Each tile hosts a distance-($d$) surface-code patch. We measure time in $t$ steps, where one $t$ corresponds to a surface code cycle. The latency of lattice surgery is also $t$ steps. For example, in practical scenarios, a $t$ time step takes around 10$\mu s$ for a $d=11$ surface code. 
We observe the design in Fig. \ref{fig:arch}. All logical data qubits are mapped to the grey outer ring; the interior white tiles form the CR and are reserved for ancilla used to measure and implement the non-Clifford gates in the workload. The MSF repeatedly distills magic states ($|m\rangle$) and supplies them to the CR. For the simplicity of the design, we consider a single, 15-to-1 distillation protocol for producing high-fidelity magic states. 
% \rev{We later discuss the magic state footprint of different sized protocols and the impact of contention in concurrent workloads.}

\subsubsection{Tile Budget}\label{subsec:budget}
We consider an \(n\times n\) logical fabric shown in Fig.~\ref{fig:arch}, excluding the
row reserved for the magic-state factory (MSF). The \(n\times n\) region above the MSF naturally
decomposes into three parts: (i) the data tiles, (ii) the ancilla tiles, and (iii) the compute region (CR).
First, the outer data ring is formed by the perimeter of the square, which contains the 
\(
\text{border tiles} = n^{2} - (n-2)^{2} = 4(n-1)
\)
tiles. We reserve one of these for the MSF channel, leaving
\(
\#\text{data tiles} = 4(n-1) - 1
\)
logical data tiles on the ring.
Second, the interior of the square comprises
\(
(n-2)^{2}
\)
ancilla tiles. The ancilla ring, given by the perimeter of the inner \((n-2)\times(n-2)\) square,
comprises
\(
\#\text{ancilla-ring tiles} = (n-2)^{2} - (n-4)^{2} = 4(n-3).
\)
of which, $4(n-3) - 1$ tiles are usable, as one of these tiles is used as the inner MSF channel tile. Finally, the remaining central tiles ($(n-4)^2$) form the CR.

\subsubsection{Movement}\label{subsubsec:movement}
We next characterise the movement and measurement cost in units of \(t\).  For data
patches on the outer ring that are not at the corners, we orient the logical \(X\) and
\(Z\) operators such that one boundary faces the adjacent ancilla ring.  A lattice-surgery
measurement of \(X\) or \(Z\) between the data patch and a neighbouring ancilla patch
then costs \(1t\).  A logical \(Y\) measurement is implemented by a short sequence of
deformations and two Pauli-product measurements; in this geometry the total cost is $5t$.
\begin{figure}    
    \centering
    \includegraphics[width=0.8\linewidth]{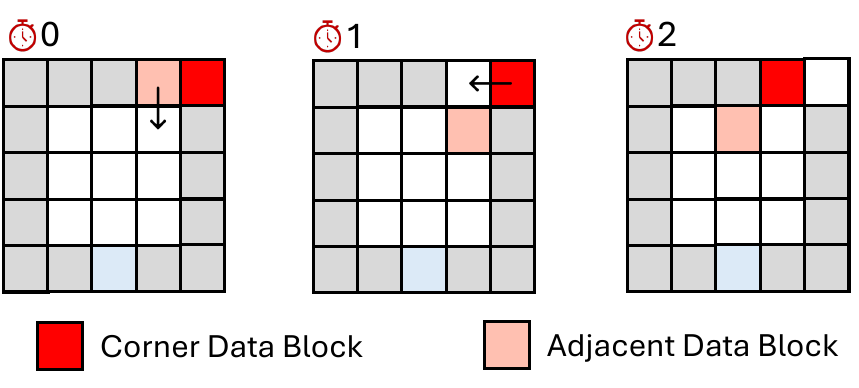}
    %\vspace{-10pt}
    \caption{\textbf{Tile movement in our proposed design.} Implementation of a corner move, where an ancilla tile is used to move the corner patch from position and expose one of the edges to the CR for $X$ and/or $Z$-basis measurement.}
    \label{fig:opscorner}
    % \vspace{-10pt}
\end{figure}
Corner patches are slightly more expensive, since they have only one neighbouring ancilla tile.  To expose the required boundary, the data patch is first shifted along
the ring or into the ancilla ring.  Shifting a corner patch to the ancilla ring costs
\(2t\) (Fig. \ref{fig:opscorner}). If the ancilla ring is occupied, moving the patch directly into the CR incurs an additional time step.  Consequently, logical \(X\) and \(Z\)
measurements for corner qubits cost at most \(3t\) (or \(4t\) in the worst case when
the CR is used), while \(Y\) measurements cost at most $7t$ (or $8t$ in the worst
case).  Importantly, these movement overheads are constant in \(n\), as every data qubit
remains at a bounded distance from the CR in the single-ring layout.

\subsection{Scaling Up the Design}
\label{subsec:entension}
\begin{figure}    
    \centering
    \includegraphics[width=1\linewidth]{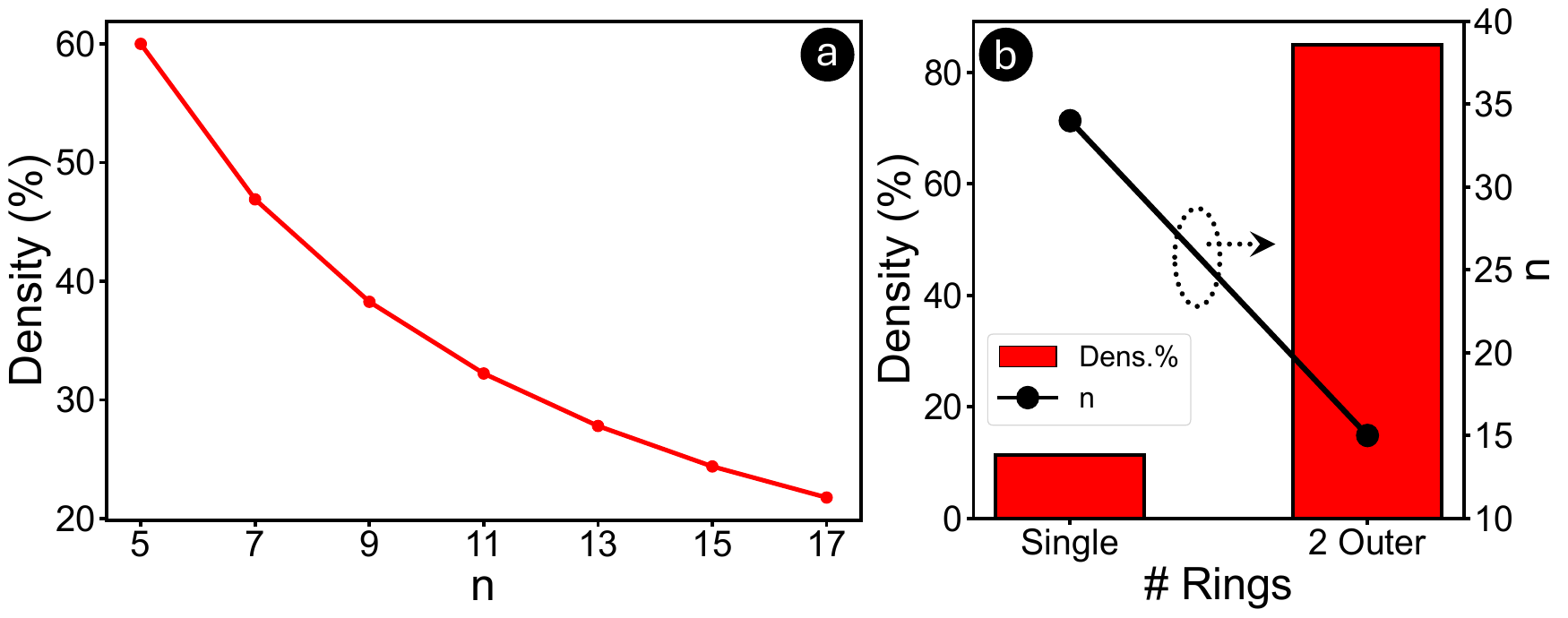}
    %\vspace{-10pt}
    \caption{\textbf{Scaling up the design.} (a) Data tile density vs size ($n\times n$ grid) of the floorplan. (b) The difference in floorplan size required to implement a QFT-128 workload, and the improvement in data tile density due to the stacking of data rings.}
    \label{fig:density}
    % \vspace{-10pt}
\end{figure}

In this sub-section, we discuss how to scale the design up for larger workloads. There are primarily two ways to scale the design up: (i) to increase $n$ for the $n\times n$ grid; and (ii) to stack tiles outside the ring to accommodate the logical qubits of the workload. We observe from Fig. \ref{fig:density}(a) that on increasing $n$, the data tile density of the layout decreases in $\Theta(1/n)$ on increasing $n$. Therefore, we follow through with the second strategy to accommodate larger workloads. From Fig. \ref{fig:density}(b), we show the improvement in the data-tile density (from $11.93\%$ to $84.7\%$) for a QFT-128 workload on stacking the outer layers while maintaining a decent size in the CR to accommodate the qubits for computation. We also end up using $\sim75\%$ lesser tiles (from a $34\times 34$ grid to a $15\times 15$ grid with two stacked data rings). However, the stacking of layers comes at the cost of an increased movement of the data tiles in the outer rings, which adds to the execution overhead minimally during the computation process. In the following portions, we discuss an optimal movement strategy, Directed Annular Movement (DAM), to reduce the movement latency as much as possible.

% \begin{figure}    
%     \centering
%     \includegraphics[width=0.4\linewidth]{samples/fig/density.png}
%     %\vspace{-10pt}
%     \caption{Plot showing the density reduction on increasing the ring size \hl{include two more plots here--mog the sweep to follow the stacking as well}}
%     \label{fig:sweep}
%     % \vspace{-10pt}
% \end{figure}
% \subsubsection{Adding layers}
% Instead, in this work we fix \(n\) (and hence the size of the compute region and ancilla
% ring) and introduce additional \emph{outer data rings} when \(w_q\) exceeds the capacity
% of the innermost ring. In this stacked-annulus layout, the innermost ring continues to
% host the “hot” data qubits, which have the highest T/Y activity and interaction degree,
% while surplus qubits are placed on one or more outer rings. This preserves a high logical
% density in the vicinity of the CR and ancilla ring, at the cost of a bounded radial
% movement overhead: qubits stored on outer rings incur an additional \(O(L)\) time to be
% moved inward, where \(L\) is the number of data rings, but the compute region and ancilla
% structure remains unchanged.

% \begin{figure}    
%     \centering
%     \includegraphics[width=0.8\linewidth]{samples/fig/extemded.png}
%     %\vspace{-10pt}
%     \caption{A diagrammatic representation of the extended architecture \hl{needs to be more descriptive and add the entry points k}}
%     \label{fig:arch}
%     % \vspace{-10pt}
% \end{figure}

% \medskip
\noindent\textbf{Directed Annular Movement (DAM):} The $T$ gates are implemented in the CR along with the magic states from the MSF. We label the inner ring of data tiles as \emph{ring} $0$, and extend by adding $L$ \emph{outer data rings}, indexed by
\(
r \in \{0,1,\dots,L\}.
\)
Ring $r$ contains $N_r$ logical tiles, indexed by an angular coordinate
\(
\theta \in \{0,1,\dots,N_r-1\}.
\)
Therefore, each surface code patch occupies exactly one tile $(r,\theta)$.
We include multiple \emph{CR-entry locations} on ring-$0$ to increase the throughput of magic states, specified by 
\(
\Theta_{\mathrm{CR}} = \bigl\{\theta_a^{(1)},\dots,\theta_a^{(K)}\bigr\} \subseteq \{0,1,\dots,N_0-1\},
\)
such that each $\theta_a^{(k)}$ has a dedicated adjacent CR slot just inside ring~$0$. 
\begin{figure}    
    \centering
    \includegraphics[width=1\linewidth]{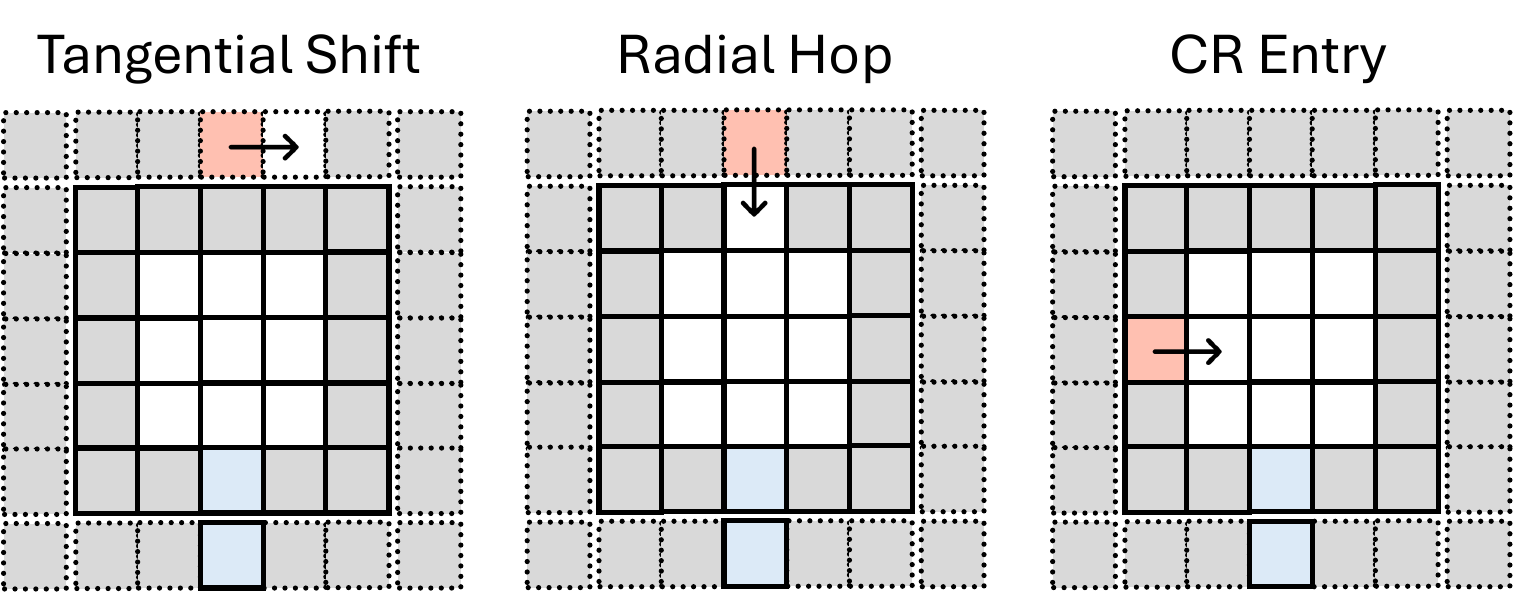}
    %\vspace{-10pt}
    \caption{\textbf{Tile movement in the design.} Three movement primitives obtained by scaling the design up (details in Section \ref{subsec:entension}).}
    \label{fig:opsextension}
    \vspace{-10pt}
\end{figure}

\noindent\textbf{Movement primitives:}
We define three movement primitives, each assuming a single $t$ step to get realized using standard surface-code patch deformations (Fig. \ref{fig:opsextension}).
\begin{enumerate}
  \item \emph{Tangential shift:} A tile moves along its ring:
  \(
  (r,\theta) \longrightarrow (r,\theta \pm 1 \bmod N_r).
  \)
  A global ring rotation is just many tangential shifts applied in parallel.

  \item \emph{Radial hop: } A tile moves one ring inward while keeping its angle:
  \(
  (r,\theta) \longrightarrow (r-1,\theta),
  \)
  provided the target tile $(r-1,\theta)$ is empty. Since the computation happens in the CR, data tiles are moved inwards only.

  \item \emph{CR entry:} A tile moves from ring~$0$ into an adjacent CR slot in \emph{constant time}:
  \(
  (0,\theta_a^{(k)}) \longrightarrow \mathrm{CR\_slot}^{(k)},
  \theta_a^{(k)} \in \Theta_{\mathrm{CR}}.
  \)
\end{enumerate}
The path of a tile is therefore completely characterized by a finite sequence of tangential shifts on its current ring, followed by a sequence of inward radial hops, followed by a single CR entry. 
% Consider a qubit $q$ initially located at tile $(r,\theta)$. The \emph{radial distance} of the tile to the inner data ring is
% \(
% d_r(q) = r,
% \)
% with $d_r(q)=0$ if $q$ already lies on ring~$0$. To define its angular distance to the nearest CR-entry lane, we let
% \(
% \mathrm{ring\_dist}_r(\theta,\theta') \;=\; \min\bigl\{ (\theta-\theta') \bmod N_r,\; (\theta'-\theta)\bmod N_r \bigr\}
% \)
% denote the wrap-around distance on ring $r$, which always satisfies
% \(
% \mathrm{ring\_dist}_r(\theta,\theta') \le \frac{N_r}{2}.
% \)
% The \emph{angular distance} of $q$ to the closest CR-entry angle on its ring is then
% \(
% d_\theta(q) = 
% \min_{\theta_a \in \Theta_{\mathrm{CR}}}
% \mathrm{ring\_dist}_r(\theta,\theta_a).
% \)
% The individual path length (in code cycles) for moving $q$ from its initial position into a CR slot is upper-bounded by
% \[
%   M(q) = d_\theta(q) + d_r(q) + c_{\mathrm{CR}},
% \]
% corresponding to first aligning $q$ with a lane using tangential shifts ($d_\theta(q)$ steps), then performing inward radial hops ($d_r(q)$ steps), and finally executing a constant-time CR entry. The asymptotic complexity of $c_{\mathrm{CR}}$ is constant in $t$, hence dropping it from the expression gives the upper bound as $M(q) = d_\theta(q) + d_r(q)$.

\noindent\textbf{Movement Latency: }
The movement latency \(T_{\mathrm{move}}(q)\) of a logical qubit \(q\), initially placed at tile \((r,\theta)\), is determined by the shortest path from its current location to a feasible interface tile adjacent to the CR. This path has two components: (i) the radial distance from the CR,
\(
d_r(q)=r,
\)
and (ii) the angular distance to the nearest feasible CR-adjacent tile on ring \(r\), denoted by \(d_\theta(q)\). For two angular positions \(\theta\) and \(\theta'\) on that ring, define the wrap-around distance as
\(
\mathrm{ring\_dist}_r(\theta,\theta')
=
\min\{(\theta-\theta') \bmod N_r,\; (\theta'-\theta)\bmod N_r\},
\)
so that \(\mathrm{ring\_dist}_r(\theta,\theta') \leq N_r/2\) (by construction). Therefore,
\(
d_\theta(q) \leq \frac{N_r}{2}.
\)
Since each tangential or radial move requires one $t$ time step, the movement latency of \(q\) is upper-bounded by
\(
T_{\mathrm{move}}(q) = d_r(q)+d_\theta(q).
\)
Substituting the bounds above gives
\(
T_{\mathrm{move}}(q)=O\!(r+N_r/2).
\)
If the architecture contains \(L\) concentric data rings, then \(r \le L-1\), and since the circumference of each ring grows linearly with \(L\), we have \(N_r=O(L)\). Therefore, the worst-case movement latency of a logical qubit is
\(
T_{\mathrm{move}}(q)=O(L).
\)

\noindent\textbf{Size of the floorplan: }
Let the CR hold at most $W_{\mathrm{CR}}(n)$ tiles concurrently, where $W_{\mathrm{CR}}$ depends on the chosen grid dimension $n$.
For a workload on \(Q\) logical qubits, let \(L_j\) denote the \(j\)-th \(T\)-layer. We define the active-qubit set of \(L_j\) as
\(
S_j = \{\, q_i \in [Q] \mid P_{j,i} \neq I \,\},
\)
where \(P_{j,i}\) denotes the Pauli operator acting on qubit \(q_i\) in layer \(L_j\),
and define the peak active-set size
\(
S_{\max}:=\max_j |S_j|.
\)
We choose $n$ such that
\(
  W_{\mathrm{CR}}(n)\ge S_{\max}, 
\)
 so CR occupancy never constrains the execution of any $T$-layer. For layers with $|S_j|<W_{\mathrm{CR}}(n)$, the unused CR slots remain idle; this does not affect the logical execution time, since the $T$-layer implementation simply occupies a subset of the available CR. The provisioned CR leaves occupancy slack that reduces routing pressure, but we do not treat this as a formal edge-disjoint routing guarantee.
 
\noindent\textbf{Calculating the time to execute a workload: }
Let $T_\mathrm{total}$ be the time to execute a workload. Each $T$-layer terminates with a single-qubit measurement in the $X$, $Y$, or $Z$ basis. Let the latency for the measurement be $t_{basis}$ (discussed in Section \ref{subsubsec:movement}). Therefore, the time for measuring a $T$-layer $L_j$ is $T_{\mathrm{meas}}(L_j)
  \;=\;
  \max_{\text{ops in } L_j} t_{\text{basis}}.$
Combining everything, a conservative estimate for the duration of a single $T$-layer $L_j$ is, therefore,
\(
  T(L_j)
  = T_{\mathrm{move}}(L_j) + T_{\mathrm{meas}}(L_j).
\)
Let $\tau_{\mathrm{MSF}}$ denote the MSF startup latency until the first usable magic state(s) arrive at the $T$-port. After this initial delay, we observe that the MSF output is uniform \cite{chatterjee2025qspellbookcraftingsurfacecode}\cite{Litinski_2019_msd}\cite{lsqca} and can be overlapped with computation such that steady-state distillation does not introduce additional stalls. Therefore, the total time to execute a workload with $L_j$ layers is
% \[
%   T_{\mathrm{total}}
%   = \sum_j T(L_j) + \tau_{MSF} =\sum_j \bigl(T_{\mathrm{move}}(L_j) + T_{\mathrm{meas}}(L_j)\bigr)+\tau_{MSF}.
% \]
\[
\begin{aligned}
T_{\mathrm{total}}
&= \sum_j T(L_j) + \tau_{\mathrm{MSF}} \\
&= \sum_j \Bigl(
    T_{\mathrm{move}}(L_j)
    + T_{\mathrm{meas}}(L_j)
   \Bigr)
   + \tau_{\mathrm{MSF}} .
\end{aligned}
\]
For example, if we choose a 15-to-1 magic state distillation protocol, it takes $11t$ to generate one high-fidelity magic state \cite{Litinski_2019}. Thus, the $T_{total}$ becomes $\sum_jT_j+11t$, as the rest of the magic states are generated by the time a workload gets executed using the existing magic state.

\section{Placement and Optimization Strategies}\label{sec:placementandoptim}

\subsection{Workload-Aware Placement}\label{sec:placement}

\begin{figure}    
    \centering
    \includegraphics[width=1\linewidth]{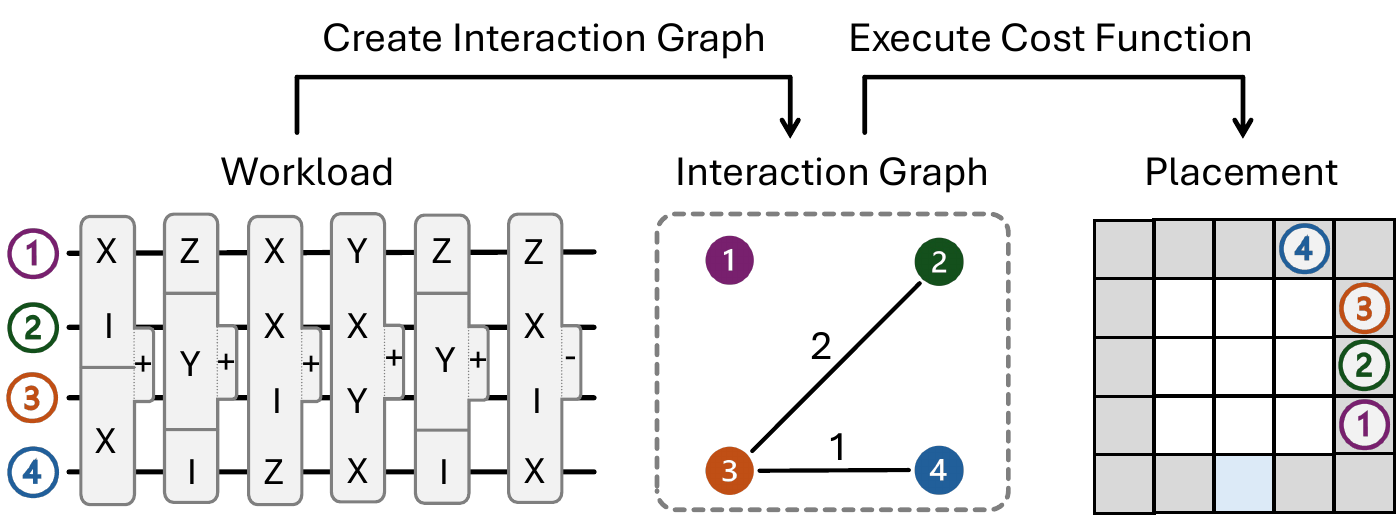}
    %\vspace{-10pt}
    \caption{\textbf{The proposed workload-aware placement policy.} The $T$ and $Y$ gate densities are computed and the multi-qubit interaction is captured into a graph (note, the weight between edges 2 and 3 is 2 due to the presence of 2 multi-qubit $T$ gates), and finally the $\mathrm{cost}(q)$ is calculated to place the logical qubits onto the floorplan.}
    \label{fig:placement}
    \vspace{-10pt}
\end{figure}

Here, we propose an efficient strategy for mapping the logical qubits to the data tiles, using the non-Clifford resource demand to guide placement onto our architecture design. $T$-gates are defined as $R_p(\pi/8)$ over the Pauli matrices $X$, $Y$, and $Z$. To design an intuitive mapping policy, the logical qubits with the most measurement-heavy $T$ gates need to be placed closest to the CR to ensure a minimal execution latency. Therefore, we consider a higher weightage of qubits with active $T$ gates and further isolate the $Y$ gates to refine our metric to place the qubits as they take more time steps in terms of $t$ to get implemented (Section \ref{subsubsec:movement}). 
For each logical qubit $q$ in a workload, we aggregate the workload characteristics into a scalar score 
\(
\mathrm{Tload}(q) := T_s(q) + \alpha_T T_m(q) ,\;
\mathrm{Yload}(q) := Y_s(q) + \alpha_Y Y_m(q),
\)
where $T_s(q)$ and $T_m(q)$ represent the number of single and multi-qubit $T$ gates on qubit $q$ respectively, $Y_s(q)$ and $Y_m(q)$ represent the number of single and multi-qubit $Y$ gates on qubit $q$ respectively, and $\alpha_T,\alpha_Y>0$ upweight multi-qubit rotations relative to single-qubit ones. We further define an interaction degree, deg$_{int}(q)$ as the total number of multi-qubit gates on the logical qubit $q$. Together, the overall cost is
\[
\mathrm{cost}(q)
= \lambda_T\,\mathrm{Tload}(q)
+ \lambda_Y\,\mathrm{Yload}(q)
+ \lambda_{\mathrm{int}}\,\deg_T(q),
\]
with $\lambda_T,\lambda_Y,\lambda_{\mathrm{int}}>0$ tunable parameters. A higher value of $\mathrm{cost}(q)$ indicates a stronger need for the qubit to be placed in the inner data rings. $\mathrm{cost}(q)$ takes into account fully the movement pressure in terms of $t$ induced by the multi-qubit $T$ gates.
After computing $\mathrm{cost}(q)$ for all
$q\in Q$, we sort the qubits in descending order of $\mathrm{cost}(q)$ and fill the data tiles starting from the innermost rings greedily: the highest-cost qubits are assigned to the innermost data ring,
followed by the next-highest to the subsequent rings, and so on.  This assigns the most resource-intensive qubits, the shortest access to the CR, while qubits with smaller non-Clifford load, and heavier identity load are placed in outer rings where movement latency is less critical. This procedure is illustrated with an example in Fig. \ref{fig:placement}. We calculate the $\mathrm{Tload}(q)$, and $\mathrm{Yload}(q)$ from the 4-qubit workload. Next, we construct the interaction graph using the multi-qubit gates to determine the deg$_{int}(q)$ and compute $\mathrm{cost}(q)$. Based on the $\mathrm{cost}(q)$ values we place the logical qubits 2 and 3 close on the edge, with 3 and 4 sharing an ancilla tile and 1 at last.

% \begin{theorem}\label{thm:movement}
% Let $\mathcal{T}$ be the set of data tiles and let $r:\mathcal{T}\to\{0,1,\dots,L\}$ denote the ring-index function.
% A placement is an injective mapping $\mathcal{P}:Q\to\mathcal{T}$.
% Let $c(0)\le c(1)\le\cdots\le c(L)$ be a nondecreasing ring penalty and let each ring $r$ have capacity $N_r$.
% Consider the placement objective
% \[
% \min_{\mathcal{P}:Q\to\mathcal{T}\ \emph{injective}} \sum_{q\in Q} \mathrm{cost}(q)\,c\bigl(r(\mathcal{P}(q))\bigr).
% \]
% Then assigning qubits in nonincreasing order of $\mathrm{cost}(q)$ to tiles in nondecreasing order of $c(r)$ achieves a global optimum.
% \end{theorem}
% We discuss the proof in \textcolor{red}{Appendix] \ref{app:proof4.1}.

\subsection{Optimizing the layout}
\label{sec:fastY}
\begin{figure}    
    \centering
    \includegraphics[width=1\linewidth]{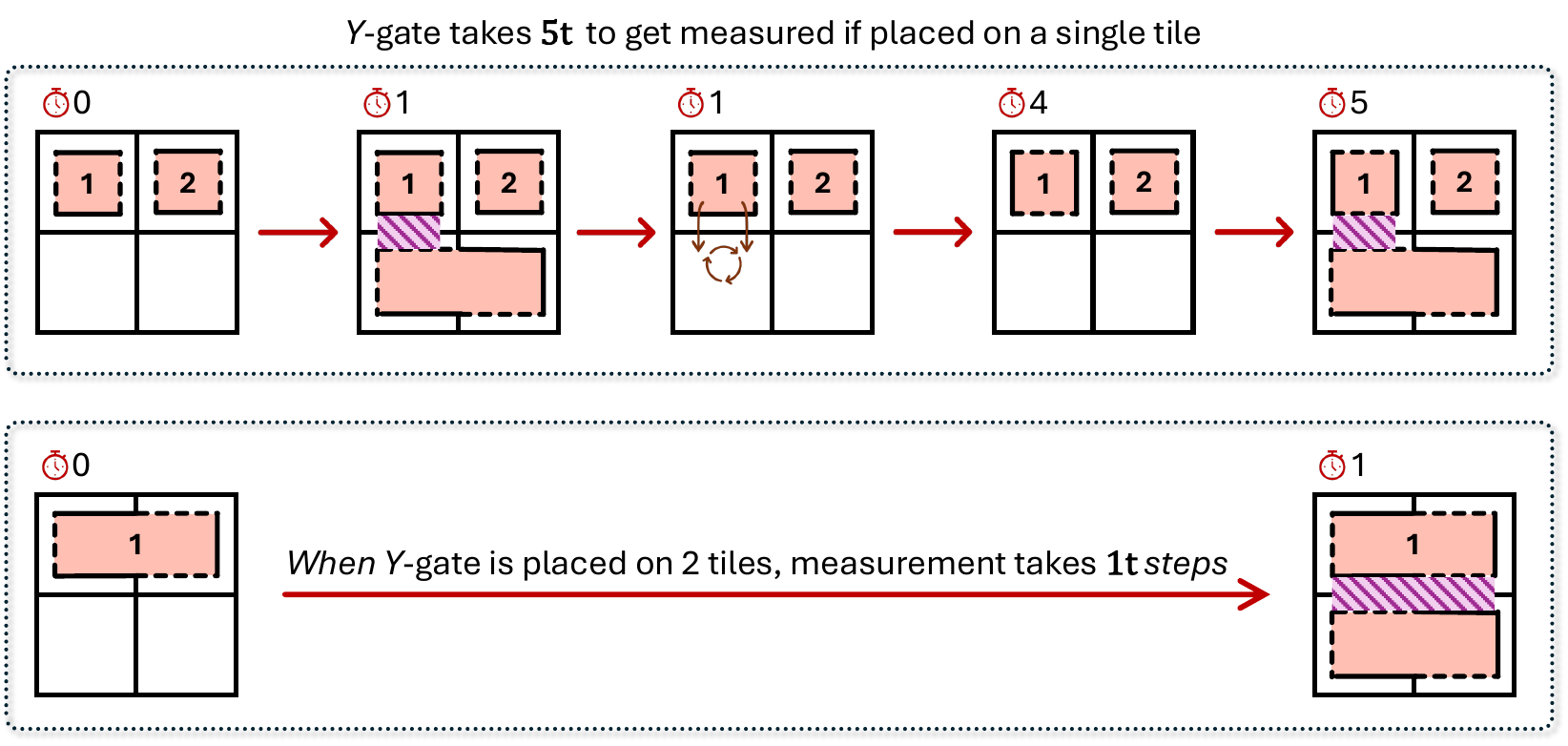}
    %\vspace{-10pt}
    \caption{\textbf{Example showing fast-$Y$ optimization.} The top block shows the un-optimized measurement taking $5t$. The bottom block shows the optimized measurement being done in $1t$ at the cost of one extra data tile.}
    \label{fig:optimise}
    \vspace{-10pt}
\end{figure}

We propose an optimization strategy to improve the performance of the design at a minimal cost of an increase in the number of tiles. In Fig. \ref{fig:optimise}  example, each logical qubit is represented by a single tile on the inner data ring in the baseline layout. Measuring a qubit in the $Y$ basis requires a sequence of rotations that takes $t_Y^{(1)}\in\{5t,7t\}$ surface-code cycles, depending on the position of the data tile in the floorplan. We introduce an optional \emph{two-tile} encoding for selected qubits, in which a logical qubit is represented by two adjacent tiles
with an edge having the $X$ and $Z$ stabilizers, enabling a $Y$ measurement in a single time step (Fig. \ref{fig:optimise}) (following the Pauli measurement idea of computing $iXZ$ to measure out $Y$). Therefore, in this encoding, the same $Y$-basis operation takes $t_Y^{(2)} = 1t$ surface code cycles, at the cost of consuming two inner-ring data tiles instead of one. 
\noindent\textbf{Initial Speedup: }
Since the two-tile assignment to a qubit will always speed up the $Y$ measurement, we can say that $t_Y^{(1)}>t_Y^{(2)}$. Therefore, for a qubit $q$, let $Y(q)$ denote the total number of single-qubit $Y$ rotations on $q$ along with the multi-qubit rotations in which the local Pauli operator on $q$ is $Y$, and we compute the per-qubit latency reduction from upgrading $q$ to a two-tile patch is $\Delta(q) := Y(q)\,\bigl(t_Y^{(1)} - t_Y^{(2)}\bigr)$. 
\noindent\textbf{Movement Penalty: }
Given the fixed inner data ring size, each two-tile upgrade therefore forces at least one other qubit to be \emph{placed} from the inner ring and relocated to an outer ring. This causes a \emph{movement penalty}. Let $I(q')$ denote the number of times qubit $q'$ must be moved if it is placed in the outer ring(s) instead of the inner ring, and $q$ can be given a two-tile upgrade. We denote $I(q') := \nu(q')\,\Delta r(q')$, where $\nu(q')$ is the number of $T$ layers in which $q'$ is active and $\Delta r(q')$ is the number of rings by which $q'$ is displaced due to eviction. This captures the fact that placing $q'$ outwards increases the number of its radial hops to the CR. Since the proximity of a qubit $q'$ placed in a data tile to the CR is determined by the $\text{cost}(q')$ (as discussed in Section \ref{sec:placement}), we also take that into consideration while determining which qubits get the two-tile benefit.
\noindent\textbf{Net tradeoff: }
Combining the initial speedup and movement penalty, we propose a simple scoring metric. Upgrading a given qubit $q$ to a two-tile patch while placing some other qubit
$q'$ from the inner ring to the outer ring yields a net latency change
\[
G(q,q') := \Delta(q) - I(q')
        = Y(q)\,\bigl(t_Y^{(1)} - t_Y^{(2)}\bigr) - I(q').
\]
We consider such an upgrade beneficial only if $G(q,q')>0$, i.e., if the
reduction in $Y$-basis time for $q$ outweighs the additional movement latency.
% incurred by placing $q'$ in an outer data ring. We precompute the initial speedup $\Delta(q)$ and movement penalty $I(q)$ for each qubit $q$, sort the qubits in $I(q)$ in the descending order of $\Delta(q)$. We greedily choose a qubit $q'$ that can be moved to the outer ring 

% We adopt a simple and efficient greedy heuristic to assign the two-tile benefit to qubits in the inner ring. For each qubit $q$ resident on the inner ring, we precompute its movement cost, $\text{cost}(q)$, the initial tradeoff $\Delta (q)$, and the movement penalty, $I(q)$.

% Let $\mathcal{I}$ denote the set of inner-ring qubits under the baseline
% (one-tile) placement. The greedy procedure then operates as follows:
% \begin{enumerate}
%   \item Sort the qubits in $\mathcal{I}$ in descending order of $\Delta(q)$.
%   \item Traverse this list; for each candidate $q$, identify the qubit
%         $q'\in\mathcal{I}$ with the smallest movement penalty $I(q')$ and smallest movement cost $\text{cost}(q')$ that can be
%         placed in the outer data ring to free up a neighbouring tile for a two-tile patch.
%   \item If $G(q,q')>0$, promote $q$ to a
%         two-tile patch and place $q'$ to the outer ring; otherwise, stop when no candidate yields positive gain.
% \end{enumerate}
% % We discuss the algorithm in Appendix \ref{app:opti-algo}.

% \input{4_Movement.tex}
\section{Sharing Resources Across Concurrent Workloads}
\label{sec:sharing}

In this section, we discuss the extension of our mapping and optimization policies on a shared floorplan among concurrent independent workloads.

\subsection{Problem Formulation and Assumptions}
\label{sec:sharing-problem}

\begin{figure}    
    \centering
    \includegraphics[width=0.5\linewidth]{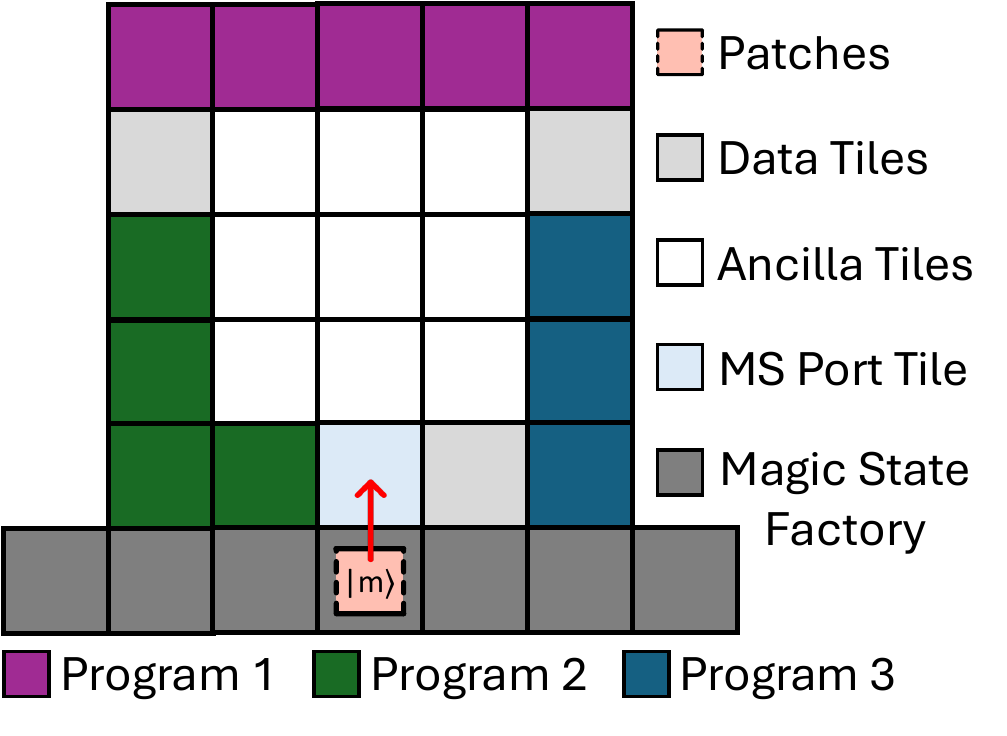}
    %\vspace{-10pt}
    \caption{\textbf{Multiprogramming in our proposed design.} A scenario of a concurrent execution of three workloads (5-qubit, 3-qubit, and 4-qubit). }
    \label{fig:multi}
    \vspace{-10pt}
\end{figure}

Let there be $W$ workloads indexed by $w\in\{1,\dots,W\}$. Workload $w$ contains a set of logical data qubits $\mathcal{Q}^{(w)}$ and is compiled into a sequence of $T$-layers
$\{L^{(w)}_j\}_{j=0}^{J_w-1}$,
with active sets
\(
S^{(w)}_j=\{q_i\in\mathcal{Q}^{(w)} \mid P_{j,i} \neq I\}.
\) where $P_{i,j}$ denotes the Pauli operator acting on qubit $q_i$ in layer $L^{(w)}_j$. The workloads share the data ring and the CR tiles.
To isolate the mapping and movement effects studied in this section, we assume the CR
and MSFs are sufficiently provisioned such that they do not impose steady-state contention under the target degree of multiprogramming. Specifically, for $W$ concurrently resident workloads, we size the CR so that
\(
  W_{\mathrm{CR}}(n)
  \;\ge\;
  \max_{\text{concurrent layers}}
  \sum_{w=1}^{W} \bigl|S^{(w)}_{j_w}\bigr|
  \;\;\;\le\;\;\;
  \sum_{w=1}^{W} S^{(w)}_{\max},
\)
ensuring that non-Clifford layers do not require batching due to CR occupancy. Likewise, we assume aggregate
MSF throughput meets peak non-Clifford demand after the initial latency of producing the magic states, so that its availability does not throttle execution. Under this assumption, multiprogramming overhead is dominated by static ring allocation and movement
contention.
The goal is to choose a full joint optimization of a combination of the aforementioned resources for all workloads. We therefore adopt a greedy approach that (i) prioritizes qubits expected to amortize inner-ring proximity and fast-$Y$ resources, (ii) limits cross-workload interference on lanes/CR-entry channels when desired, and (iii) yields a tractable execution-time estimate by extending our per-layer time model.

\subsection{Greedy Pooling via Per-Workload Costs}
\label{sec:multiworkload_cost_qgp}

We reuse the single-workload placement features from Section \ref{sec:placement} to quantify how strongly each
qubit benefits from proximity to the CR. For each workload $w$, we compute a per-qubit
\(
  cost^{(w)}(q)
  =
  \lambda_T\,Tload^{(w)}(q)
  +
  \lambda_Y\,Yload^{(w)}(q)
  +
  \lambda_{\mathrm{int}}\,deg_T^{(w)}(q)
\)
where $  q\in \mathcal{Q}^{(w)}$, $Tload^{(w)}(q)$ and $Yload^{(w)}(q)$ aggregate the one- and multi-qubit non-Clifford
structure involving $q$, and $deg_T^{(w)}(q)$ captures the weighted non-Clifford interaction degree.
A higher value of $cost^{(w)}(q)$ indicates a stronger need for $q$ to reside in inner rings where the radial distance to the CR is smaller.

In a multi-workload setting, na\"ively placing the global top-$N_0$ qubits (by $cost^{(w)}(q)$)
onto ring~0 can yield high throughput but poor isolation: a single workload can monopolize ring~0,
lanes, and fast-$Y$ opportunities. We therefore introduce a lightweight quota stage followed by
greedy placement within each quota. We define three scalar pressures per workload: (i) The aggregate active sets per workload $P_T(w)
  := \sum_{j=0}^{J_w-1} \bigl|S^{(w)}_j\bigr|$; (ii) The aggregate $Y$ gates on qubit $q$ in the workload $w$, $P_Y(w) := \sum_{q\in Q^{(w)}} Y^{(w)}(q)$; and (iii) the total movement latency in the workload $w$, as defined in Section \ref{subsec:entension}, $T_\mathrm{move}(w)
  := \sum_{q\in Q^{(w)}} \Bigl(d_r(q) + d_\theta(q)\Bigr)$. For the allocation of the qubits on the tiles, we determine a selection rule by sorting the qubits by $\text{cost}^{(w)}(q)$ and then selecting the top few qubits using a combination of the workload pressures defined earlier. We define an inner-data ring tile budget $  B_0(w)
  = \left\lceil
  N_0 \cdot
  \frac{\eta_T P_T(w) + \eta_M P_M(w)}
       {\sum_{w'} \eta_T P_T(w') + \eta_M P_M(w')}
  \right\rceil, $ and a fast-$Y$ budget $B_Y(w)
  = \left\lceil
  B_Y^{\mathrm{tot}} \cdot
  \frac{P_Y(w)}{\sum_{w'} P_Y(w')}
  \right\rceil$ with tunable parameters $\eta_T,\eta_M\ge 0$ and a system-level cap $B_Y^{tot}$ which is determined by the ring 0 capacity of the entire architecture.

Within each workload $w$, we sort qubits by $cost^{(w)}(q)$ and select the top $B_0(w)$ to reside
on ring~0. Remaining qubits are placed on outer rings via the same greedy radial fill used in
Section~4 (highest-cost qubits on smaller radii). This selection is affected by:
(1) non-Clifford activity and interaction structure through $cost^{(w)}(q)$, and
(2) workload pressure via $B_0(w)$, which allocates inner-ring real estate to workloads expected to
amortize it through either non-Clifford throughput or reduced movement.

\subsection{Fast-$Y$ Allocation Under Multiprogramming}
\label{sec:fasty_multi}

We reuse the single-workload fast-$Y$ model of the baseline architecture in Section \ref{sec:fastY} in the multi-workload setting, with one
additional restriction for isolation: fast-$Y$ promotions and the resulting movements are performed
\emph{within} each workload $w$ (i.e., a promotion of a workload-$w$ qubit may only move another
workload-$w$ qubit). Each workload receives a budget $B_Y(w)$ promotions.

For workload $w$, the benefit of promoting a ring~0 qubit $q$ is
\(
  \Delta^{(w)}(q) := Y^{(w)}(q)(t_Y^{(1)}-t_Y^{(2)}),
\)
and the movement penalty for a displaced neighbor $q'$ is
\(
  I^{(w)}(q') := \nu^{(w)}(q')\,\Delta r(q').
\)
We apply the same greedy rule as in Section \ref{sec:fastY}: we process candidate $q$ in descending order of
$\Delta^{(w)}(q)$; for each $q$, we choose an adjacent data tile
\(
  q^\star \in \arg\min_{q'} \bigl(I^{(w)}(q'),\,cost^{(w)}(q')\bigr),
\)
and perform the promotion if $\Delta^{(w)}(q)-I^{(w)}(q^\star)>0$, stopping after $B_Y(w)$ successful
promotions or when no positive-gain move remains.

\section{Evaluation}\label{sec:eval}

\noindent\textbf{Benchmarks:}
We evaluate our proposed architecture design using a three-part benchmark suite: (i) The first part comprises electronic structure Hamiltonians of the molecules, H$_2$, LiH, and H$_2$O. We considered a single step of a Trotter decomposition for these circuits; (ii) The second part is composed of the arithmetic circuits, viz., Quantum Fourier Transform, adder, multiplier, and the square root circuit; and (iii) The third part is a synthetic set of Clifford+$T$ circuits with randomised size and $T$-gate density to stress-test the efficacy of the proposed design. This is done to include a set of circuits in out test benchmark that have a higher circuit depth and density than the standard circuits (first two parts of our suits). We generate a pool of 1000 circuits with the number of qubits and $T$ layers varying from 10 to 1000, and variable $T$ gate densities that divides the total circuit set into three classes, low density ($1\%-\sim 30\%$), medium density ($\sim30\%-\sim 60\%$), and high density ($\ge60\%$).

\noindent\textbf{Experimental Setup:} All algorithms were implemented in Python 3.12.1 on an Intel Core i9-13900K CPU with a clock frequency of 4.8GHz and a 16GB of RAM. The hyperparameters for the algorithms were chosen carefully after sweeping through the possible cases. Through ablation we find $\alpha_Y/\alpha_T=1.5$; $\lambda_T:\lambda_Y:\lambda_{\mathrm{int}}=1:2:4$; and $\eta_M/\eta_T = 2$. A detailed discussion is present in Section \ref{sec:ablation}.

\noindent\textbf{Evaluation Metrics:}
For a proper evaluation of our proposed architecture design, we report two normalized metrics, viz., the cycles per $T$-instruction ($CPI_T$), and the routing inflation ratio ($\rho_{route}$). 
% From Section \ref{subsec:entension}, we express the total runtime of a workload with $J$ $T$-layers as 
% \(
% T_{\mathrm{total}} =\sum_{j=1}^J\Big(T_{\mathrm{move}}(L_j)+T_{\mathrm{meas}}(L_j)\Big)
% +\tau_{\mathrm{MSF}}.
% \)
We further define the total number of non-Clifford instructions as $N_T =\sum_{j=1}^J\lvert S_j\rvert$, where $S_j$ is the active set of $\pi/8$ Pauli-product rotations in each $T$-layer $L_j$. Therefore, 
\[CPI_T := \frac{T_{total}}{N_T}.\] CPI$_T$ measures the average number of code cycles expended per $\pi/8$ rotation. Therefore, a reduction in $CPI_T$ means the architecture and mapping strategy convert each unit of non-Clifford
progress into fewer surface-code time steps. This normalization enables fair comparison across
benchmarks with different sizes and $T$-counts.
%We additionally decompose $CPI_T$ into additive components to obtain $CPI_{move}=\frac{\sum_{j=1}^J T_{\mathrm{move}}(L_j)}{N_T}$ and $CPI_{meas}=\frac{\sum_{j=1}^J T_{\mathrm{meas}}(L_j)}{N_T}$. We can drop the contribution for the rest of the components as they are bounded in our architecture design. 
While $CPI_T$ captures efficiency of placement, we also isolate the \emph{architecture-induced movement latency}. To capture the movement overhead specifically, we define an inflation factor relative to movement, as 
\[
\rho_{\mathrm{route}}
\;:=\;
1 \;+\; \frac{\sum_{j=1}^{J} T_{\mathrm{move}}(L_j)}{\sum_{j=1}^{J} T_{\mathrm{meas}}(L_j)}.
\]
Here $\rho_{\mathrm{route}} \approx 1$ indicates that movement overhead is negligible relative to measurement latency, whereas larger values indicate that movement and contention dominate runtime.

Through our evaluation we attempt to answer the following \emph{research questions}:
\begin{itemize}
    \item [\textbf{RQ1: }]How does our proposed architecture design perform and compare to prior work?
    \item [\textbf{RQ2: }]What amount of performance benefit is obtained from the fast-$Y$ optimization knob?
    \item [\textbf{RQ3: }]Does workload-aware placement matter, and how robust is it to parameter choices?
    \item [\textbf{RQ4: }]How does the design perform on concurrent execution of workloads?
    \item [\textbf{RQ5:}]How does our architecture perform with respect to the magic state contention?
\end{itemize}

\subsection{(RQ1) Performance analysis}

\begin{figure*}    
    \centering
    \includegraphics[width=1\linewidth]{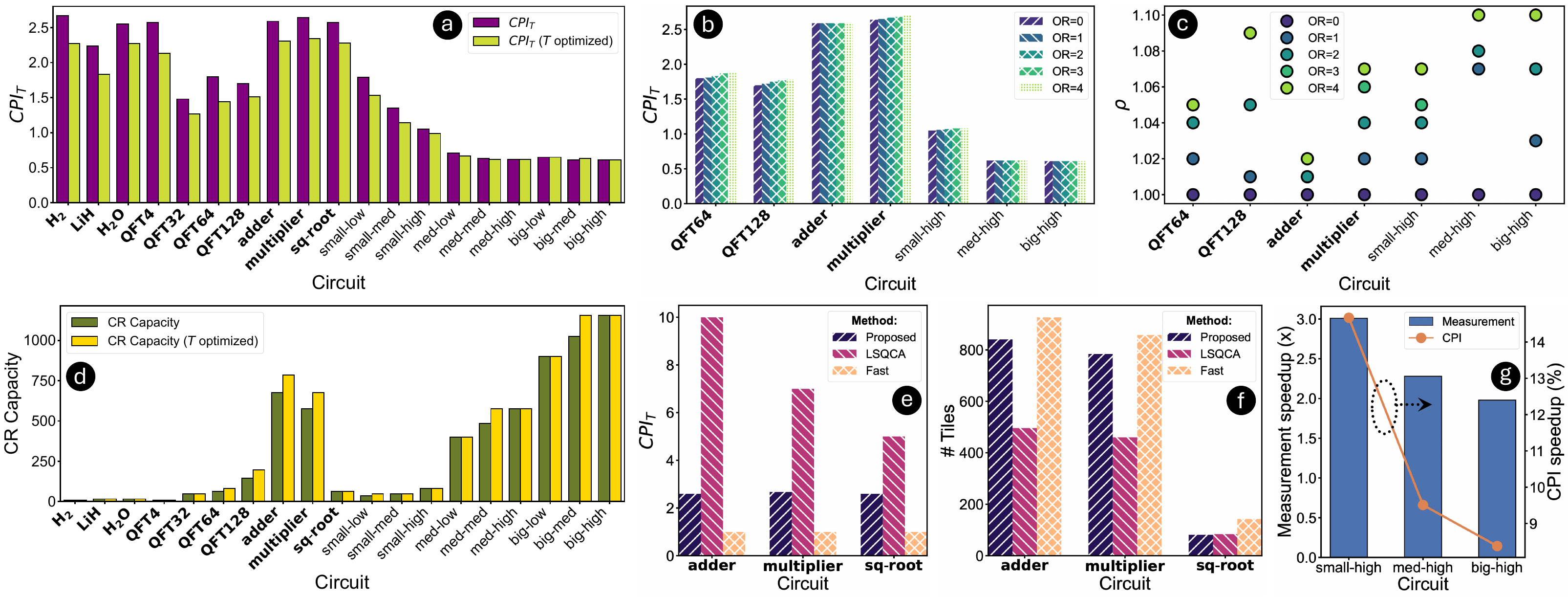}
    %\vspace{-10pt}
    \caption{\textbf{Plots representing RQ1 and RQ2.} (a) Spread of $CPI_T$ across the benchmark circuits. (b) Negligible change in $CPI_T$ with more outer rings (OR) in the proposed design. (c) Change in the routing inflation factor $\rho_\mathrm{route}$ with more outer rings (OR) in the proposed design. (d) shows the CR Capacity for the benchmark circuits. (e) and (f) Improvement over both $CPI_T$ and data tile density in the proposed design compared to the state-of-the-art \cite{lsqca}\cite{Litinski_2019}. (g) Improvement in $T_{\mathrm{meas}}$ and $CPI_T$ with the fast-$Y$ optimization. Benchmark labels shown in bold correspond to quantum circuits for real-world problems, whereas non-bold labels denote synthetic workloads generated with varying sizes and $T$-gate densities.}
    \label{fig:rq12}
    % \vspace{-10pt}
\end{figure*}

We evaluate single-program execution to quantify (i) end-to-end efficiency per non-Clifford instruction and (ii) the movement overhead induced by routing under our annular mapping. We report the two metrics, viz., $CIP_T$, and $\rho_{route}$ as discussed before. 
\noindent\textbf{Benchmark-wide efficiency across workloads: } 
Fig. \ref{fig:rq12}(a) reports $CPI_T$ over our benchmark suite, spanning the Hamiltonian instances (e.g., H$_2$, LiH, H$_2$O), structured transforms (QFT family), arithmetic workloads (adder, multiplier, square-root), and the synthetic workloads. For the experiments, we consider the fast-$Y$ optimization set to on. The results exhibit a broad spread in $CPI_T$, indicating that efficiency of execution is workload-dependent and largely governed by non-Clifford \emph{amortization}. We observe that, in smaller circuits with a lower $T$-gate density, the movement and phase operation latencies constitute a larger fraction of $T_{\mathrm{total}}$, inflating $CPI_T$. Conversely, workloads with denser non-Clifford structure amortize these fixed costs over a larger $N_T$, yielding lower $CPI_T$. The intermediate behavior of the QFT family is consistent with its regular structure: it incurs repeated routing patterns, but maintains sufficient $T$-density to avoid the high-overhead in $CPI_T$ observed in the sparsest circuits. We also report the CR capacity required for each workload in Fig.~\ref{fig:rq12}(d). The CR is provisioned according to the peak active-set size across all $T$-layers, i.e., we choose $W_{\mathrm{CR}}(n)\geq S_{\max}$. Therefore, workloads with wider non-Clifford layers require larger CR regions, while narrower workloads can be supported with smaller CRs. This result contextualizes the $CPI_T$ values in Fig.~\ref{fig:rq12}(a): the reported execution efficiency is obtained while explicitly provisioning enough CR occupancy for the simultaneously active qubits in each $T$-layer.

\noindent\textbf{Architectural scaling: }
To test robustness under architectural scaling, we sweep the number of outer rings (represented as OR in Fig. \ref{fig:rq12}(b), with OR=0 meaning the design only comprises the inner ring) and track both $CPI_T$ and $\rho_{route}$ for representative workloads (Fig. ~\ref{fig:rq12}(b) and ~\ref{fig:rq12}(c)). Similar to the previous experiments, we keep the fast-$Y$ optimization set to on with four radial lanes to CR entry when the number of outer rings increases. Across the larger QFT instances and a subset of the custom workloads (the ones with the highest $T$-gate densities), $CPI_T$ remains nearly unchanged as OR increases, indicating that moderate changes in radial depth do not materially degrade end-to-end efficiency. In contrast, $\rho_{route}$ increases monotonically with OR, as expected: additional rings increase the maximum radial displacement required before qubits reach ring 0 resources. Importantly, the observed inflation remains modest over the sweep, demonstrating that perimeter scaling introduces a bounded movement penalty rather than a collapse in throughput. This behavior is consistent with the one-way DAM mechanism: each data qubit moves radially inward at most once, so added perimeter capacity primarily affects a bounded prefix of the path length rather than inducing repeated long-range paths.

\noindent\textbf{Effect of $T$-depth optimizers:}
We further explore the impact of $T$-parallelism on our architecture. On using a $T$-depth optimizer on the compilation pipeline, the number of sequential non-Clifford operations is reduced by exposing the inherent $T$-gate parallelism in the workload. This reduces the $T$-depth and overall $T$-count, increasing the peak active-set size of the workload ($S_{max}$). This impacts the CR provisioning per workload. To evaluate this, we run the $T$-depth optimization pipeline in \cite{Amy_2014} (under the 0-ancilla constraint), and report the $CPI_T$ and CR capacity values in Fig. \ref{fig:rq12}(a) and Fig. \ref{fig:rq12}(d), respectively. From Fig. \ref{fig:rq12}(d), we observe that the optimized circuits require equal or larger CR sizes to accommodate the increased $T$-gate density. This increase is stepwise following the discrete function defined in Section \ref{subsec:budget}. There are cases where there is no increase in the CR region, stating that the provisioned CR has sufficient occupancy slack. From Fig. \ref{fig:rq12}(c), we observe a dip in the $CPI_T$ of $\sim 15\%$ in the quantum chemistry workloads, $\sim11\%$ in the arithmetic circuits, and $\sim16\%$ in the QFT implementations. For our synthetic benchmarks, the $CPI_T$ remains unchanged for the larger workloads, with a $\sim 13\%$ decrease in the $CPI_T$ for the smaller workloads. This behavior indicates that by increasing the $T$-gate parallelism in the workloads, we can obtain a lower $CPI_T$ for sparser and smaller workloads due to the increase in the corresponding $T$-gate density, but the $CPI_T$ remains unchanged for denser workloads due to the original placement already optimizing for the lowest possible movement and placement latency.

\noindent\textbf{Comparison to prior work:}
We compare our architecture design to the current state-of-the-art LSQCA \cite{lsqca} and the fast block design \cite{Litinski_2019}.
For LSQCA, we model the cost of moving logical qubits between the dense memory region and the compute region through explicit load/store operations. For the \emph{Fast} block design, we model its support for unit-time $Y$-basis measurements, while also accounting for the larger logical tile footprint required by this capability. This ensures an apples-to-apples comparison between the methods.
The design of LSQCA has been proposed to increase the memory density of the floorplan to reduce the qubit overhead, whereas the fast block design measures any $T$-gate in $1t$ irrespective of its Pauli operator. We observe the comparison of these designs to our proposed baseline and our optimal variation of the same in Fig. \ref {fig:rq12}(e) and \ref {fig:rq12}(f). LSQCA exhibits substantially higher $CPI_T$ on these circuits, reflecting the additional routing and movement overheads in the dense memory region. The fast block shows a $CPI_T$ of 1 as expected. Our baseline improves over the LSQCA $CPI_T$ by $3.86\times$, $2.61\times$, and $1.93\times$, and our optimal variation uses $9.3\%$, $8.6\%$, and $43.4\%$ less tiles than the \emph{Fast} block design for the adder ($433q$), multiplier ($400q$), and square root ($60q$) workloads, respectively. For a $d=11$ surface code implementation, this corresponds to a savings of $\sim20,000$, $\sim17,000$, and $\sim15,000$ qubits for the adder ($433q$), multiplier ($400q$), and square root ($60q$) workloads, respectively, over the \emph{Fast} block design. These results validate our claim that the proposed design explores both axes of design, viz., the execution time and the resource overhead, showing the improvement in data tile density (less requirement of resources) with an improved execution time (lesser $CPI_T$). Specifically, the relatively higher percentage of qubit reduction for the square root workload demonstrates the efficacy of our proposed design for near-term fault-tolerant circuits. We further observe that the \emph{Fast} block design \cite{Litinski_2019} is prone to resource contention when all data patches are used to abstract logical qubits, which degrades the $CPI_T$ of the workload and can increase it beyond 1.

\noindent\textbf{Scalability: }
As discussed in Section \ref{sec:placement}, under the DAM in our annular design, the displacement of any qubit is bounded by
\(
T_\mathrm{move}= d_\theta(q) + d_r(q) = O(L),
\)
where $d_\theta(q)$ is the angular distance to its assigned entry lane, $d_r(q)$ is the radial distance of a qubit placed in a tile in the outer rings, and $L$ is the number of data rings. In contrast to the denser design proposals, such as LSQCA, the worst-case movement scales in terms of $O(\sqrt n)$, where $n>>L$. Therefore, in our case, we find that the cost of movement is amortized in larger workloads, leading to a minimal increase in $\rho_{route}$ (by $\sim10^{-1}$) and $CPI_T$ ($\sim10^{-2}$).

\rqsummarybox{\textbf{RQ1 Summary: }Larger workloads and $T$-dense workloads have lower $CPI_T$ due to the amortization benefit of measurement over the minimal movement. The optimized design in our architecture keeps the same $CPI_T$ by using up to less than $\sim 
21\%$ data tiles.}

\subsection{(RQ2) Optimization gains}

We next isolate the contribution of our fast-$Y$ capability, which accelerates the measurement of $Y$ components in $\pi/8$ Pauli-product rotations by promoting a selected subset of data qubits to a fast-$Y$ measurement path. Since this optimization only affects the measurement term in our runtime model, it provides a controlled way to quantify how much of the end-to-end runtime is attributable to measurement overhead (as opposed to routing and conversion). For these experiments, we consider the number of outer rings to be set to four. Concretely, under the decomposition of our $T_{\mathrm{total}}$, therefore, enabling fast-$Y$ directly reduces $T_{\mathrm{meas}}(L_j)$ for those qubits whose measurement basis includes a $Y$ component, while leaving $T_{\mathrm{move}}(L_j)$ unchanged.
To ensure this ablation is informative, we evaluate on a subset of our custom dataset of circuits with the highest $T$-gate density, and consequently the highest $Y$-gate count, making $T_{\mathrm{meas}}$ a meaningful bottleneck and allowing the fast-$Y$ knob to be stress-tested. This choice avoids the trivial regime where $Y$ measurements are rare and $T_{\mathrm{meas}}$ is a negligible fraction of $T_{\mathrm{total}}$, in which case fast-$Y$ would have little observable impact on either runtime or $CPI_T$.
Figure~\ref{fig:rq12}(g) reports two outcomes: (i) the \emph{measurement speedup}, defined as
\(
S_{\mathrm{meas}} \;:=\; T_{\mathrm{meas}}^{base}/T_{\mathrm{meas}}^{(\mathrm{fast}\text{-}Y)},
\)
and (ii) the \emph{$CPI_T$ improvement}, reported as a percentage reduction
\(
\Delta CPI_T(\%) \;:=\; (CPI_T^{base}-CPI_T^{(\mathrm{fast}\text{-}Y)})/CPI_T^{base}.
\)
Across the workloads, enabling fast-$Y$ yields a clear multiplicative reduction in $T_{\mathrm{meas}}$ (left axis), confirming that the optimization is effective when $Y$-basis measurements are frequent. We find an average of $\sim2.42\times$ improvement in the $T_{\mathrm{meas}}$, evaluated over the total of 1000 workloads. 
However, the corresponding reduction in $CPI_T$ (right axis) is necessarily more modest, because $CPI_T$ aggregates \emph{all} per-layer costs. We observe an average reduction of $10.86\%$ in the $CPI_T$ of all workloads. 
The figure also shows that the realized $T_{\mathrm{meas}}$ speedup decreases from the smallest to the largest circuit family. This trend is consistent with two effects that become more pronounced at scale: (i) non-measurement costs ($T_{\mathrm{move}}$ and $T_{\mathrm{T}}$) constitute a larger fraction of runtime as the active set of qubits increase, and (ii) the fast-$Y$ capability is applied to a bounded subset of qubits at any time, so the fraction of measurements that benefit from the accelerated path can decrease as the workload grows. Consequently, while fast-$Y$ consistently improves the measurement subsystem, its impact on $CPI_T$ exhibits diminishing returns as other costs dominate the critical path in larger workloads.
\rqsummarybox{\textbf{RQ2 Summary: } Enabling the fast-$Y$ optimization knob delivers a $\sim2.42\times$ average $T_{\mathrm{meas}}$ speedup and $\sim 10.86\%$ $CPI_T$ reduction at a minimal cost of data tiles. }

\subsection{(RQ3) Ablation Study}\label{sec:ablation}

% \begin{figure*}    
%     \centering
%     \includegraphics[width=1\linewidth]{samples/fig/rq_3_plots.pdf}
%     %\vspace{-10pt}
%     \caption{Plots representing RQ3. }
%     \label{fig:rq3}
%     % \vspace{-10pt}
% \end{figure*}

\begin{figure*}    
    \centering
    \includegraphics[width=1\linewidth]{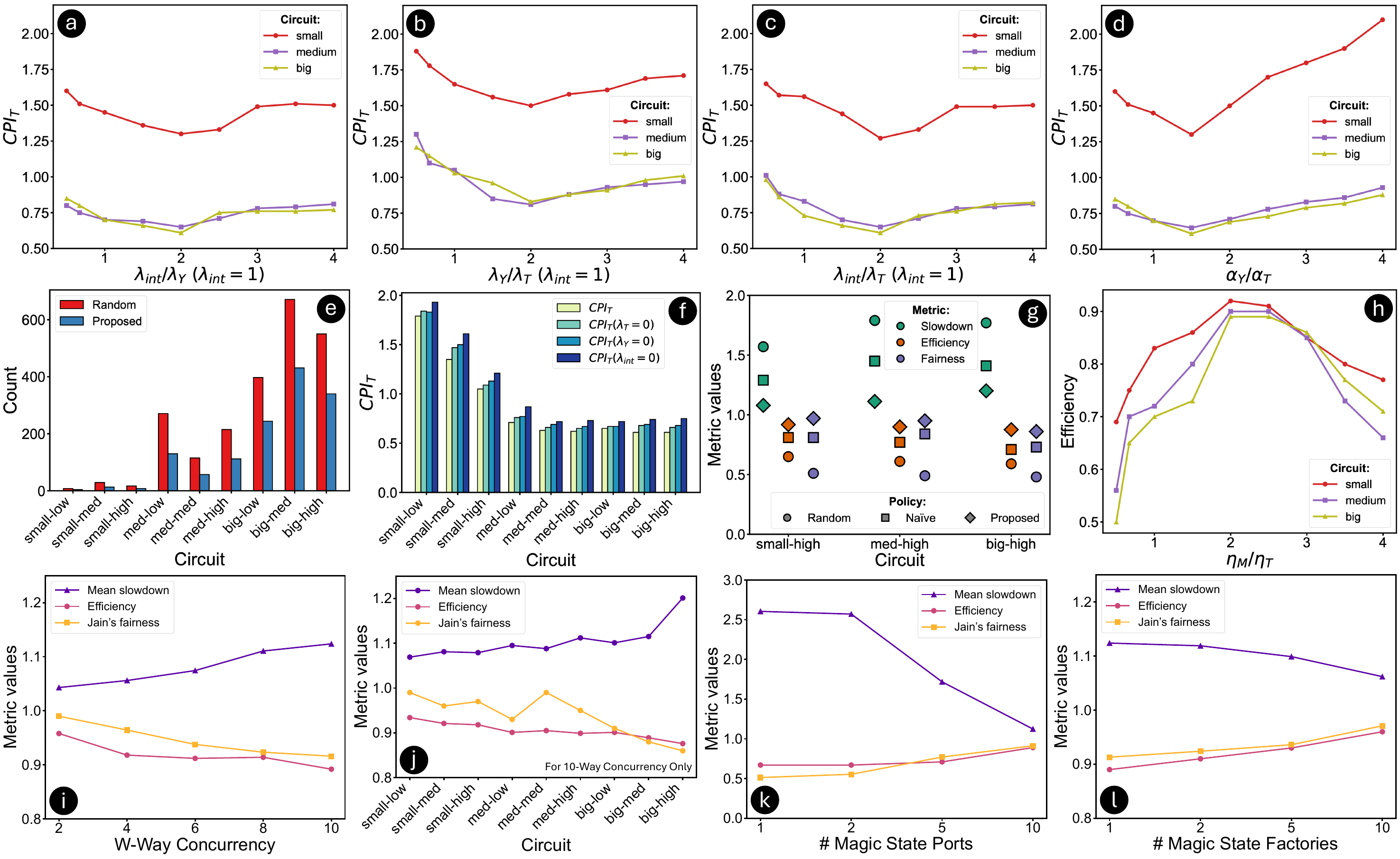}
    %\vspace{-10pt}
    \caption{\textbf{Plots representing RQ3, RQ4, and RQ5.} (a)-(d) Ablation over the hyperparameters of $\mathrm{cost}_q$. (e) Ablation on the movement in the architecture design by comparing the proposed policy with a random one. (f) Ablation over the $\mathrm{cost}(q)$ to show the sensitivity of the performance to the individual components of the design. (g) Ablation on the performance over high $T$-gate density workloads. (h) Scaling of $\eta_M$ and $\eta_T$ with respect to efficiency. (i) Sweep across different concurrency levels. (j) Detailed analysis for different workload mixes for a 10-way concurrency. (k) Contention of magic state ports under a single 15-to-1 distillation factory. (l) Increase in throughput on increasing factories at 10 ports.}
    \label{fig:rq345}
    \vspace{-10pt}
\end{figure*}

In this subsection, we conduct an ablation study to attribute performance gains to two key components of our placement policy: (i) the workload-awareness of our proposed placement policy reduces the movement latency toward the CR, and (ii) the composite qubit-scoring function $\mathrm{cost}(q)$ that prioritizes which qubits should reside closer to ring~0/CR access. Since our design executes measurements within the CR, and the non-Clifford execution phase operation latency $T_T(\cdot)$, and the measurement latency, $T_{\mathrm{meas}}$, is unaffected by where a qubit initially resides, the primary placement-sensitive term in the runtime decomposition is the movement component. Specifically, in $T_{\mathrm{total}}$, the placement policy only changes the cost of reaching CR resources (captured by $T_{\mathrm{move}}$). Therefore, to isolate the effect of placement alone, we report the movement-only outcomes in the first ablation. For the experiments, we set the fast-$Y$ optimization to on and the number of outer rings and radial lanes to four.

\noindent\textbf{Ablation 1 (Movement-only): }
Figure~\ref{fig:rq345}(e) compares the total movement latency (in terms of $t$) for a random placement baseline versus our proposed placement across the synthetic workload families. The proposed policy consistently reduces movement across all circuit classes, with the gap widening as workload size increases. As is expected, a random placement policy places $T$-heavy and $T$-light qubits with equal likelihood without optimizing for the distance from the CR, thereby increasing angular travel and delaying admission into the CR. In contrast, our placement explicitly aligns qubits to lanes and prioritizes ring~0 residency for those with higher expected non-Clifford participation and temporal urgency, thereby reducing the aggregate ingress distance. We observe an average of $\sim2.42\times$ decrease for the smaller circuits, $\sim2\times$ decrease for medium-sized circuits, and $\sim1.71\times$ decrease for the larger circuits in the $T_{\mathrm{move}}$ latency.

\noindent\textbf{Ablation 2: The cost of placement ($\mathrm{cost}(q)$): }
We next evaluate the sensitivity of performance to the individual components of the qubit-scoring function, $\mathrm{cost}(q)$. We perform a controlled ablation by setting individual weights $\lambda_i$, for $i \in \{T,Y,\mathrm{int}\}$ to zero, thereby removing that term from $\mathrm{cost}(q)$ while leaving all other mechanisms unchanged. Figure~\ref{fig:rq345}(f) reports the resulting $CPI_T$ for each workload under the full model and under each ablated variant (e.g., $CPI_T(\lambda_T=0)$, $CPI_T(\lambda_Y=0)$, and $CPI_T(\lambda_{\mathrm{int}}=0)$). We find that the $CPI_T$ changes measurably, by an average of $\sim8.6\%$ for smaller workloads, $\sim18.2\%$ for medium and larger workloads, by turning $\lambda_{\mathrm{int}}$ off. On turning $\lambda_{\mathrm{T}}$, and $\lambda_{\mathrm{Y}}$ off we observe an average increase in $CPI_T$ by $\sim7\%$ in both cases, confirming that all the individual components are important in the $\mathrm{cost}(q)$ cost function. For the smaller workloads, we observe a slightly lower degradation on turning off $\lambda_{\mathrm{int}}$ than the medium and larger workloads, which is expected due to the presence of a larger number of multi-qubit gates in them than in the smaller workloads. 

\noindent\textbf{Ablation 3: Hyperparameter sweep: } To obtain the hyperparameter values for our algorithms, we conduct a sweep over the dimensionless ratios of the hyperparameters. We perform the experiments over our custom benchmark set with the highest $T$-gate densities with varying qubit counts (small, medium, big). Fig. \ref{fig:rq345}(d) represents the relation between the hyperparameters $\alpha_Y$ and $\alpha_T$. We obtain a value of 1.5 as ratio when the minimum $CPI_T$ obtained with the value increasing thereafter. This shows the preference towards $Y$-critical qubits outweighs the other $T$-gates and increases the movement latency when the ratio is higher, hence choosing a moderate value of 1.5, which lies in the minimum, is stable for the algorithm. Fig. \ref{fig:rq345}(a), \ref{fig:rq345}(b), and \ref{fig:rq345}(c) sweeps the relative importance of multi-qubit interaction (represented by $\lambda_\mathrm{int}$). We obtain a ratio of $\lambda_T:\lambda_Y:\lambda_{\mathrm{int}}=1:2:4$ when the $CPI_T$ is lowest across all benchmarks, showing that the interaction degree is the dominant factor in determining the placement of logical qubits on the data tiles efficiently. In Fig. \ref{fig:rq345}(h), we find the sweep of the relative weight of the movement pressure ($P_M$) and the non-Clifford gate pressure ($P_T$) in the ring 0 partition rule. We find that increasing the ratio $\eta_M/\eta_T$ improves concurrent utilization up to a value of 2, after which the efficiency of execution of the workloads drops.

\rqsummarybox{\textbf{RQ3 Summary: } We find that our proposed placement policy gives a \emph{twice} benefit on the $T_{\mathrm{move}}$ times over different-sized workloads, with the multi-qubit interaction degree affecting the $CPI_T$ most (increases by $\sim18\%$).}

\subsection{(RQ4) Multiprogramming}\label{sec:multiresult}

% \begin{figure}    
%     \centering
%     \includegraphics[width=0.7\linewidth]{samples/fig/rq_4_plots.pdf}
%     %\vspace{-10pt}
%     \caption{Plots representing RQ 4. We report the metrics: mean slowdown, efficiency, and Jain's fairness index for the executed workloads. In (a) we present the metrics over our custom benchmark set. In (b) we ablate the performance over high $T$-gate density workloads.}
%     \label{fig:rq4}
%     % \vspace{-10pt}
% \end{figure}

We next evaluate our multi-workload mapping policy under concurrent execution, where multiple workloads share ring~0 capacity and the CR admission bandwidth. In this setting, the objective is no longer single-program efficiency alone, but a three-way tradeoff between (i) overall throughput, (ii) per-workload performance isolation, and (iii) fairness of service. To capture these aspects in a workload-agnostic manner, we report the \emph{mean slowdown}, \emph{efficiency}, and \emph{Jain's fairness index} for every workload.

\noindent\textbf{Metrics: }
For a set of $W$ workloads executed concurrently, let $T_w^{\mathrm{alone}}$ denote the runtime of workload $w$ when executed in isolation on the same architecture, and let $T_w^{\mathrm{conc}}$ denote the completion time of workload $w$ under concurrency. We define the per-workload slowdown as
\(
\mathrm{SD}_w :=\frac{T_w^{\mathrm{conc}}}{T_w^{\mathrm{alone}}}\;\;\ge\; 1,
\)
and report the \emph{mean slowdown} as $\frac{1}{W}\sum_{w=1}^{W}\mathrm{SD}_w$. To quantify aggregate utilization, we report an \emph{efficiency} metric (a normalized throughput proxy) defined as the ratio of total useful isolated work to the concurrent makespan:
\(
\mathrm{Efficiency}:= \frac{\sum_{w=1}^{W} T_w^{\mathrm{alone}}}{W\cdot T_{\max}^{\mathrm{conc}}}, T_{\max}^{\mathrm{conc}} := \max_{w} T_w^{\mathrm{conc}}.
\)
This quantity lies in $(0,1]$, with higher values indicating better overlap and less wasted time due to contention and head-of-line blocking. Finally, we quantify fairness using Jain's fairness index over the achieved per-workload service rates. Using normalized service $x_w := 1/\mathrm{SD}_w$ (larger is better), Jain's fairness index is
\(
\mathcal{J} := \frac{\left(\sum_{w=1}^{W} x_w\right)^2}{W\cdot \sum_{w=1}^{W} x_w^2}, 0<\mathcal{J}\le 1,
\)
where $\mathcal{J}=1$ indicates perfectly equalized slowdowns across workloads, while smaller values indicate that some workloads are disproportionately penalized under sharing. For all experiments, we consider ten workloads executed concurrently with fast-$Y$ optimization set to on, and each workload has its own separate CR entry channel. We also assume that resources are sufficient and workloads arrive offline, as our goal is to explore the efficiency of placement and mapping using our strategy.

\noindent\textbf{Sensitivity in Concurrency:} We conduct a sweep over multiple levels of concurrency (Fig. \ref{fig:rq345}(i)), using a mixed pool of workloads drawn from all circuit classes in our benchmark suite, averaged over five independent random samples. We observe a monotonic degradation in the efficiency and fairness, while the slowdown increases as we enable resource sharing between an increasing number of workloads. The mean slowdown increases from $1.044$ for 2 workloads to $1.124$ for 10 workloads, suggesting an increase of only $\sim8\%$ despite increasing the workload count by 5 times. Similarly, efficiency drops by $\sim7\%$, while fairness remains above $\sim0.9$. We can infer from these results that no single workload is disproportionately penalized under high contention.

\noindent\textbf{Scaling in the proposed sharing policy:}
Figure~\ref{fig:rq345}(j) reports the metrics under the proposed multi-programming policy for 10 concurrent workloads. As the workload size and pressure increase (moving from \emph{small} to \emph{big}, and from \emph{low} to \emph{high}), the mean slowdown increases, reflecting the fact that larger and denser workloads place greater demand on shared ring~0 residency. We find an average \emph{mean slowdown} $1.1$ across all workloads. Correspondingly, \emph{efficiency} drops from an average of $\sim 92.4\%$ in smaller workloads to $\sim90.1\%$ in medium-sized and $\sim88.6\%$ in larger workloads, indicating that the concurrent system becomes increasingly dominated by shared-resource contention. Jain's index also trends downward, from $\sim0.99$ in smaller workloads with a sparser $T$-gate density to $\sim0.85$ in lager workloads with the highest $T$-gate density, which is consistent with increased heterogeneity in which workloads as they contend for the critical path at different times; under high contention, slight asymmetries in arrival patterns or $T$-layer structure can amplify into unequal service, reducing fairness.

\noindent\textbf{Ablation across policies: }
To attribute these multi-programming trends to the policy choices, Figure~\ref{fig:rq345}(g) compares three placement policies on representative workload instances with the highest $T$-gate densities: \emph{Random}, \emph{Na\"ive}, and \emph{Proposed}. The random policy assigns qubits completely randomly without workload and $T$-gate awareness, and the na\"ive policy keeps the same level of randomness but introduces the $T$-gate awareness (through $\mathrm{cost}(q)$). We find an average \emph{mean slowdown} of $\sim1.71$ and $\sim1.38$ for random and na\"ive policies, respectively, which is significantly higher, as compared to $\sim 1.1$ obtained for our proposed policy. This is expected as both random and na\"ive policies increase the angular travel of the qubits due to random placement leading to the higher mean slowdown. The efficiency also drops to an average of $\sim61.6\%$ and $\sim76.3\%$ for the random and na\"ive policies, respectively. We also observe a drop in Jain's fairness index to $\sim0.49$ for random and $\sim0.79$ for the na\"ive policy. The trend of na\"ive being slightly better than random but overall worse than our proposed policy is expected due to the introduction of partial structure in assigning qubits based on $\mathrm{cost}(q)$. In contrast, the proposed policy handles contention efficiently by partitioning the ring~0 residency and fast-$Y$ allocations to qubits per workload based on the $T$ and $Y$ gate density, therefore, increasing the efficiency and fairness in allocation and decreasing the overall slowdown in workload execution.
\rqsummarybox{\textbf{RQ4 Summary:} Under 10-way concurrency, the proposed sharing policy slows the system down by $\sim$1.1$\times$ with efficiency maintained at $\sim90\%$.}

\subsection{(RQ5) Magic State Contention}\label{subsec:msf}

% \begin{figure*}    
%     \centering
%     \includegraphics[width=1\linewidth]{samples/fig/rq_5_concurrency.pdf}
%     %\vspace{-10pt}
%     \caption{RQ 5, increase font, shaped pointers}
%     \label{fig:rq5}
%     % \vspace{-10pt}
% \end{figure*}

We evaluate the impact of magic state provisioning on our architecture model in this section and address the single workload and concurrent execution scenarios separately.

\noindent\textbf{Single Workload:} For a single workload, the MSF assumption is benign in steady state. Once the initial distillation delay $\tau_{\text{MSF}}$ has passed, the factory produces one high-fidelity magic state per $t$ in steady state \cite{Litinski_2019_msd}\cite{Litinski_2019}. We confirm this by evaluating the $CPI_T$ for $T$-dense workloads under the 15-to-1 ($\tau_{\text{MSF}}=11t$), 225-to-1 ($\tau_{\text{MSF}}=15t$), 20-to-4 ($\tau_{\text{MSF}}=17t$), and the 116-to-12 ($\tau_{\text{MSF}}=99t$). In all cases, the difference in $CPI_T$ is negligible (of the order $10^{-4}$). Since the movement latency dominates in larger circuits with higher $T$-depth and $T$-count, this is an expected outcome, and the startup cost for the magic state factories is amortized over the $T$-layers, justifying our assumption for single-workload execution. 

\noindent\textbf{Concurrent Workloads:} For execution of concurrent workloads, we study the contention of resources in Fig.\ref{fig:rq345}(k)(l), under a 10-way concurrency, using a 15-to-1 magic state protocol, since it takes the least amount of tiles to get implemented (11 tiles). To work around magic-state port contention, we design a simple priority queue with the priority order based on the wait time for a workload (workloads with higher stall time are addressed first). We further evaluate under two settings: (i) we observe the port contention by increasing the number of available ports in the architecture with a single factory producing magic states (Fig. \ref{fig:rq345}(k)), and (ii) we fix 10 ports for 10 workloads to eliminate contention and observe any additional performance gains on additional throughput by having multiple factories (Fig. \ref{fig:rq345}(l)). From Fig. \ref{fig:rq345}(k), we find that the slowdown improves from $\sim2.6$ to $\sim1.1$ while the efficiency increases by $\sim24\%$ and the fairness improves by $\sim44\%$, when the number of ports is increased from 1 in the entire architecture to 10 ports (eliminating contention). When we fix 10 ports and increase the additional throughput by having more factories (Fig. \ref{fig:rq345}(l)), we find that there is a slight monotonic improvement in all metrics (slowdown improves by $\sim5\%$, efficiency by $\sim7\%$, and fairness by $\sim6.5\%$). However, the practicality of increasing the number of factories is not justified, as the performance gains are negligible while the tile cost continues to grow linearly.

\rqsummarybox{\textbf{RQ5 Summary:} Under a 10-way concurrency, increasing magic state ports reduces the mean slowdown of the system by $\sim2.3\times$, and increasing the number of factories provides minimal gains at an increase in the tile count.}

\section{Related Works and Discussion}\label{sec:related}

Prior work on patch-based surface-code architectures has largely explored the FTQC floorplanning design space from the perspective of \emph{throughput-oriented} layouts. The tile-based framework of Litinski \cite{Litinski_2019,Litinski_2019_msd} makes explicit the trade-off between tile footprint and non-Clifford throughput via compact, intermediate, and fast data-block organizations. Compact layouts minimize area but may require multiple time steps for magic-state consumption, whereas fast layouts allocate additional ancilla space to support near unit-time execution of non-Clifford operations. More broadly, several FTQC floorplans reserve substantial auxiliary area for routing and lattice-surgery access in order to provide low-latency interaction among logical qubits \cite{Beverland_2022,Chamberland2022TwistFreeTemporallyEncoded,beverland2022assessingrequirementsscalepractical,Lee2021TensorHypercontraction}.
A contrasting line of work emphasizes \emph{density-oriented} organization. LSQCA \cite{lsqca}, for example, separates a small computational region from a denser logical memory and transfers qubits between them through load/store-style operations. This design improves storage density and avoids keeping all qubits in an operation-ready state, but it introduces variable access latency that can accumulate under repeated use. Beyond floorplan geometry, other recent work has also studied lattice-surgery scheduling and dependency-aware mapping. These approaches formulate execution as routing and ordering problems using abstractions such as 3D path routing, dependency graphs, and edge-disjoint paths \cite{hamada2026efficienthighperformanceroutinglatticesurgery,dascot,silva,Beverland2022EdgeDisjointPaths}. SWIPER \cite{swiper} accelerates time-sensitive execution through speculative decoder scheduling, \cite{ftqc_sync} introduces the idea of synchronization of syndrome-generation cycles, and \cite{ravi2022betterworstcasedecodingquantum} introduces lightweight common-case decoding together with bandwidth allocation and stall management. In our work, we address the goal of saving space and reducing execution time by implementing an efficient placement idea and a movement strategy. While these works are orthogonal to our research direction, we view these techniques as composable with our design. In parallel, direct studies of concurrent execution on shared FTQC fabrics remain limited, with recent work only beginning to address settings such as online job arrival and scheduling for fault-tolerant multiprogramming \cite{Wakizaka_2025}. 

\textbf{Future Work:} We assume a Pauli-product measurement model, which is natural for lattice-surgery execution. Here the Clifford operations are propagated by updating Pauli bases and are absorbed into terminal measurements, while the scheduled non-Clifford operations are represented as $\pi/8$ Pauli-product rotations. Our runtime model preserves intra-layer parallelism by executing active sets of each $T$-layer in the compute region and charging the layer by the maximum measurement latency among active operations. We do not evaluate separate compilation flows like the $CNOT+H,S,T$ scheduling pipeline. Extending the placement objective with a $ CNOT$ gate  interaction term is an important direction for future work. 
Our proposed architecture is not expected to alter the logical error rates or decoder-level thresholds, which have been characterized in prior work \cite{sparo}\cite{hetec}. However, systematically studying its impact on larger-scale quantum chemistry and optimization workloads, as well as evaluating its implementation within surface-code simulators, remains an important direction for future work.
Other directions may include the exploration of implementing edge-disjoint paths for the CR region to evaluate the routing pressure and design algorithms to optimize that.

\section{Conclusion}\label{sec:conc}

% In this paper, we present an annular, reconfigurable surface-code FTQC architecture that concentrates logical qubits near a central ancilla zone while preserving routing slack in outer rings. We propose (i) a workload-aware greedy placement policy, (ii) fast-$Y$ optimization, and (iii) a multiprogramming policy with fairness metrics. Our numerical evaluation presents significant gains over the current SOTA.

We presented a workload-aware, annular surface-code architecture
that jointly balances logical-qubit density and execution latency.
The proposed design organizes data patches in stacked rings around
a central compute region and uses directed annular movement to
provide bounded access to lattice-surgery resources. We further
introduced a workload-aware placement policy that prioritizes
qubits according to their non-Clifford activity and interaction
degree, together with a selective fast-$Y$ optimization that
allocates additional tiles only when the resulting measurement
speedup outweighs the movement penalty.

Across quantum-chemistry, arithmetic, QFT, and synthetic
benchmarks, the proposed architecture maintains near-optimal
execution efficiency while using up to approximately $21\%$ fewer
data tiles. The fast-$Y$ optimization provides an average
$2.42\times$ reduction in measurement latency and a $10.86\%$
reduction in $\mathrm{CPI}_T$. Under 10-way multiprogramming, the
sharing policy maintains approximately $90\%$ efficiency with an
average slowdown of about $1.1\times$. These results demonstrate
that workload-aware floorplanning can provide a practical
space--time tradeoff for scalable and reconfigurable FTQC systems.

\section{Acknowledgements}
\noindent A.G. and S.G.'s work, and A.C.'s work partly, is supported in part by the National Science Foundation (NSF) (CNS-1722557, CCF-1718474) and gifts from Intel. The remainder of A.C.’s contribution was supported in part by the U.S. Department of Energy, Office of Science, National Quantum Information Science Research Centers, Quantum Systems Accelerator (Award No. DE-SCL0000121), under Contract No. DE-AC02-05CH11231, and partly by the National Energy Research Scientific Computing Center (NERSC), a U.S. Department of Energy Office of Science User Facility (Project No. m888).

\bibliography{refs}
\bibliographystyle{ieeetr}
\end{document}